\documentstyle[12pt,epsfig]{article}
\input epsf

\begin{document}
\baselineskip=20.5pt

\def\beqra{\begin{eqnarray}} \def\eeqra{\end{eqnarray}}
\def\beqast{\begin{eqnarray*}} \def\eeqast{\end{eqnarray*}}
\def\beq{\begin{equation}}      \def\eeq{\end{equation}}
\def\be{\begin{enumerate}}   \def\ee{\end{enumerate}}

\def\fnote#1#2{\begingroup\def\thefootnote{#1}\footnote{#2}\addtocounter
{footnote}{-1}\endgroup}

\def\itp#1#2{\hfill{NSF-ITP-{#1}-{#2}}}

\def\bet{\beta}
\def\gam{\gamma}
\def\Gam{\Gamma}
\def\la{\lambda}
\def\eps{\epsilon}
\def\La{\Lambda}
\def\si{\sigma}
\def\Si{\Sigma}
\def\al{\alpha}
\def\Tha{\Theta}
\def\tha{\theta}
\def\vphi{\varphi}
\def\del{\delta}
\def\Del{\Delta}
\def\ab{\alpha\beta}
\def\om{\omega}
\def\Om{\Omega}
\def\mn{\mu\nu}
\def\mun{^{\mu}{}_{\nu}}
\def\kap{\kappa}
\def\rsi{\rho\sigma}
\def\beal{\beta\alpha}

\def\til{\tilde}
\def\rta{\rightarrow}
\def\eqv{\equiv}
\def\nab{\nabla}
\def\pa{\partial}
\def\sit{\tilde\sigma}
\def\ul{\underline}
\def\indt{\parindent2.5em}
\def\nd{\noindent}

\def\rsi{\rho\sigma}
\def\beal{\beta\alpha}

\def\caa{{\cal A}}
\def\cb{{\cal B}}
\def\cac{{\cal C}}
\def\cd{{\cal D}}
\def\ce{{\cal E}}
\def\cf{{\cal F}}
\def\cg{{\cal G}}
\def\ch{{\cal H}}
\def\ci{{\cal I}}
\def\cj{{\cal{J}}}
\def\ck{{\cal K}}
\def\cl{{\cal L}}
\def\cm{{\cal M}}
\def\cn{{\cal N}}
\def\cO{{\cal O}}
\def\cp{{\cal P}}
\def\car{{\cal R}}
\def\cs{{\cal S}}
\def\ct{{\cal{T}}}
\def\cu{{\cal{U}}}
\def\cv{{\cal{V}}}
\def\cw{{\cal{W}}}
\def\cx{{\cal{X}}}
\def\cy{{\cal{Y}}}
\def\cz{{\cal{Z}}}

\def\cdgc{C^{\dagger}C}
\def\ccdg{CC^{\dagger}}
\def\pcdgc{C'^{\dagger}C'}
\def\pccdg{C'C'^{\dagger}}
\def\vdgv{v^{\dagger}v}
\def\gmn{\cg^{\mu}_\nu}
\def\gmm{\cg^{\mu}_\mu}
\def\gmnb{\cg^{~\,\mu}_{0\,\nu}}
\def\smn{\Sigma^{\mu}_\nu}
\def\pdgp{\phi^{\dagger}\phi}
\def\ga{\cg^{(a)}}
\def\gamn{\cg^{(a)\mu}_{~~\nu}}
\def\gamm{\cg^{(a)\mu}_{~~\mu}}
\def\sia{\Si^{(a)}}
\def\siamn{\Si^{(a)\mu}_{~~\nu}}
\def\Gak{\Gam^{(a)}_{2k}}
\def\invp{\cg_0^{-1}}

\def\raisenot{\raise .5mm\hbox{/}}
\def\nota{\ \hbox{{$a$}\kern-.49em\hbox{/}}}
\def\notA{\hbox{{$A$}\kern-.54em\hbox{\raisenot}}}
\def\notb{\ \hbox{{$b$}\kern-.47em\hbox{/}}}
\def\notB{\ \hbox{{$B$}\kern-.60em\hbox{\raisenot}}}
\def\notc{\ \hbox{{$c$}\kern-.45em\hbox{/}}}
\def\notd{\ \hbox{{$d$}\kern-.53em\hbox{/}}}
\def\notbd{\ \hbox{{$D$}\kern-.61em\hbox{\raisenot}}} 
\def\note{\ \hbox{{$e$}\kern-.47em\hbox{/}}}
\def\notk{\ \hbox{{$k$}\kern-.51em\hbox{/}}}
\def\notp{\ \hbox{{$p$}\kern-.43em\hbox{/}}}
\def\notq{\ \hbox{{$q$}\kern-.47em\hbox{/}}}
\def\notW{\ \hbox{{$W$}\kern-.75em\hbox{\raisenot}}}
\def\notz{\ \hbox{{$Z$}\kern-.61em\hbox{\raisenot}}}
\def\notpa{\hbox{{$\partial$}\kern-.54em\hbox{\raisenot}}}

\def\fo{\hbox{{1}\kern-.25em\hbox{l}}}  
\def\rf#1{$^{#1}$}
\def\bx{\Box}
\def\tr{{\rm Tr}}
\def\rmtr{{\rm tr}}
\def\dgg{\dagger}
\def\trm{{\rm tr}_{_{\left(M\right)}}~}
\def\trn{{\rm tr}_{_{\left(N\right)}}~}
\def\trnn{{\rm tr}_{_{\left(N+1\right)}}~}
\def\tr2n{{\rm tr}_{_{\left(2N\right)}}~}

\def\lag{\langle}
\def\rag{\rangle}
\def\bmid{\big|}

\def\vlap{\overrightarrow{\La p}} 
\def\lrta{\longrightarrow} \def\lrar{\raisebox{.8ex}{$\longrightarrow$}}
\def\rlarw{\longleftarrow\!\!\!\!\!\!\!\!\!\!\!\lrar}
\def\vx{\vec x}  
\def\vy{\vec y}  

\def\llra{\relbar\joinrel\longrightarrow}           
\def\mapright#1{\smash{\mathop{\llra}\limits_{#1}}} 
\def\mapup#1{\smash{\mathop{\llra}\limits^{#1}}} 

\def\nmasymptotic{
{_{\displaystyle{\rm lim}}\atop
{\scriptstyle N,M\rightarrow\infty}
}\,\, 
}

\def\nasymptotic{{_{\stackrel{\displaystyle\longrightarrow}
{N\rightarrow\infty}}\,\, }} 
\def\masymptotic{{_{\stackrel{\displaystyle\longrightarrow}
{M\rightarrow\infty}}\,\, }} 
\def\wasymptotic{{_{\stackrel{\displaystyle\longrightarrow}
{w\rightarrow\infty}}\,\, }} 

\def\asymptext{\raisebox{.6ex}{${_{\stackrel{\displaystyle\longrightarro
w}{x\rightarrow\pm\infty}}\,\, }$}} 
\def\epsilim{{_{\textstyle{\rm lim}}\atop_{\epsilon\rightarrow 0+}\,\, }} 

\def\7#1#2{\mathop{\null#2}\limits^{#1}}        
\def\5#1#2{\mathop{\null#2}\limits_{#1}}        
\def\too#1{\stackrel{#1}{\to}}
\def\tooo#1{\stackrel{#1}{\longleftarrow}}
\def\nout{{\rm in \atop out}}

\def\one{\raisebox{.5ex}{1}}
\def\BM#1{\mbox{\boldmath{$#1$}}}

\def\ltsim{\matrix{<\cr\noalign{\vskip-7pt}\sim\cr}}
\def\gtsim{\matrix{>\cr\noalign{\vskip-7pt}\sim\cr}}
\def\haf{\frac{1}{2}}


\def\place#1#2#3{\vbox to0pt{\kern-\parskip\kern-7pt
                             \kern-#2truein\hbox{\kern#1truein #3}
                             \vss}\nointerlineskip}

\def\illustration #1 by #2 (#3){\vbox to #2{\hrule width #1 height 0pt 
depth
0pt
                                       \vfill\special{illustration #3}}}

\def\scaledillustration #1 by #2 (#3 scaled #4){{\dimen0=#1 \dimen1=#2
           \divide\dimen0 by 1000 \multiply\dimen0 by #4
            \divide\dimen1 by 1000 \multiply\dimen1 by #4
            \illustration \dimen0 by \dimen1 (#3 scaled #4)}}

\def\ON{{\cal O}(N)}
\def\UN{{\cal U}(N)}
\def\bdPh{\mbox{\boldmath{$\dot{\!\Phi}$}}}
\def\bPh{\mbox{\boldmath{$\Phi$}}}
\def\bPhs{\bPh^2}
\def\sef{S_{eff}[\tilde\rho]}
\def\sigx{\sigma(x)}
\def\pix{\pi(x)}
\def\bph{\mbox{\boldmath{$\phi$}}}
\def\bphs{\bph^2}
\def\ex{\BM{x}}
\def\exs{\ex^2}
\def\xdot{\dot{\!\ex}}
\def\y{\BM{y}}
\def\ys{\y^2}
\def\ydot{\dot{\!\y}}
\def\pat{\pa_t}
\def\pax{\pa_x}
\def\cia{C_{i\alpha}}
\def\cjb{C_{j\beta}}
\def\Gz{G(z)}
\def\log{{\rm log}~}
\def\Re{{\rm Re}~}
\def\Im{{\rm Im}~}
\def\nh{{\rm non-hermitean matrix}}
\def\det{{\rm det}~}
\renewcommand{\thesection}{\arabic{section}}
\renewcommand{\theequation}{\thesection.\arabic{equation}}

\itp{01}{24}

\vspace*{.0in}
\begin{center}
{\large\bf ``SINGLE RING THEOREM" AND THE DISK-ANNULUS PHASE 
TRANSITION}\end{center}
\vspace{-.2in}
\begin{center}
{\bf Joshua Feinberg$^{a *}$, R. Scalettar$^{b *}$ \& A. 
Zee$^{c }$\fnote{*}{{\it e-mail addresses: joshua@physics.technion.ac.il, 
rst@solid.ucdavis.edu, zee@itp.ucsb.edu}}}
\end{center}
\begin{center}
$^{a)}${Physics Department,}\\
{University of Haifa at Oranim, Tivon 36006, Israel\fnote{**}{{\it permanent address}}}\\
{and}\\
{Physics Department,}\\
{Technion, Israel Institute of Technology, Haifa 32000, Israel}\\
$^{b)}${Physics Department,}\\
{University of California, Davis, CA 95616, USA}\\
$^{c)}${Institute for Theoretical Physics}\\
{University of California\\ Santa Barbara, CA 93106, USA}\\
\end{center}
\begin{minipage}{6.1in}
{\abstract
Recently, an analytic method was developed to study 
in the large $N$ limit non-hermitean random matrices that are drawn from a 
large class of circularly symmetric non-Gaussian probability distributions, 
thus extending the existing Gaussian non-hermitean literature. One 
obtains an explicit algebraic equation for the integrated density of 
eigenvalues from which the Green's function and averaged density of 
eigenvalues could be calculated in a simple manner. Thus, that formalism may 
be thought of as the non-hermitean analog of the method due to Br\'ezin, 
Itzykson, Parisi and Zuber for analyzing hermitean non-Gaussian random 
matrices. A somewhat surprising result is the so called 
``Single Ring" theorem, namely, that the domain of the eigenvalue 
distribution in the complex plane is either a disk or an annulus. 
In this paper we extend previous results and provide simple new explicit 
expressions for the radii of the eigenvalue 
distiobution and for the value of the eigenvalue density at the edges of 
the eigenvalue distribution of the non-hermitean matrix in terms of moments of the eigenvalue 
distribution of the associated hermitean matrix. We then present several numerical verifications of the 
previously obtained analytic results for the quartic ensemble and its phase transition from a 
disk shaped eigenvalue distribution to an annular distribution. Finally, 
we demonstrate numerically the ``Single Ring" theorem for the sextic potential, namely, 
the potential of lowest degree for which the ``Single Ring" theorem has non-trivial consequences.}
\end{minipage}

\vspace{10pt}
\vfill
\pagebreak

\setcounter{page}{1}

\section{Introduction}
\setcounter{equation}{0}

There has been considerable interest in random non-hermitean matrices 
in recent years. Possible applications range over several areas of
physics\cite{nonherm, stony, fz, fz1}. For some recent reviews 
see\cite{reviews}.
One difficulty is that the eigenvalues of non-hermitean matrices invade the 
complex plane, and consequently, various methods developed over the years to 
deal with random hermitean matrices are no longer applicable, as these methods
typically all involve exploiting the powerful constraints of analytic function
theory. (See in particular the paper by Br\'ezin, Itzykson, Parisi, and Zuber
\cite{BIPZ}.) In \cite{fz}, two of us proposed a ``method of hermitization", 
whereby a problem involving random non-hermitean matrices can 
be reduced to a problem involving random hermitean matrices, to which various 
standard methods (such as the diagrammatic method\cite{bzw}, or the 
``renormalization group" method\cite{french, rg, daz, rectangles}) can be 
applied. An idea similar to the ``method of hermitization" was expressed 
independently in \cite{stony}.

To our knowledge, the literature on random non-hermitean 
matrices\cite{nonherm, stony} has focussed 
exclusively on Gaussian randomness. For instance, it has been known for over 
thirty years, from the work of Ginibre\cite{ginibre}, that for the Gaussian 
probability distribution $P(\phi) =(1/Z) {\rm exp}~(-N\rmtr\phi^\dgg\phi)$ 
(here, as in the rest of this paper, $\phi$ denotes an $N\times N$ complex 
random 
matrix with the limit $N\rightarrow\infty$ understood), the density of 
eigenvalues of $\phi$ is uniformly distributed over a disk of radius 1 in the 
complex plane. 

Analytic determination of the density of eigenvalues of a 
non-Gaussian probability distribution of the form 
\beq\label{prob}
P(\phi) = {1\over Z} e^{-N\rmtr V(\phi^\dgg\phi)}\,,
\eeq  
where $V$ is an arbitrary polynomial of its argument, was given for the first
time in \cite{fz1}. Based on the method of hermitization, it was shown in 
\cite{fz1} that by a simple trick, the desired density of 
eigenvalues could be obtained with a minimal amount of work, by judiciously 
exploiting the existing literature on random hermitean matrices. 

Due to the symmetry of $P(\phi)$ under the transformation 
$ \phi\rightarrow e^{i\alpha}\phi$, the density of eigenvalues is obviously 
rotational invariant. It was shown in \cite{fz1} that the class of 
probability distributions of the form (\ref{prob}) exhibits a universal 
behavior in the sense that whatever the polynomial $V$ was, the shape of 
the eigenvalue distribution in the complex plane was always either a disk 
or an annulus. This result was referred to in \cite{fz1} as the ``Single 
Ring Theorem". 

In a certain sense, the formalism developed in \cite{fz1} may be 
thought of as the analog of the work of Br\'ezin et al. for random hermitean 
matrices \cite{BIPZ}; they showed how the density of eigenvalues of hermitean 
matrices $\varphi$ taken from the probability distribution 
$P(\varphi) = (1/Z){\rm exp} [-N\rmtr V(\varphi)]$ 
with $V$ an arbitrary polynomial can be determined, and not 
just for the Gaussian case studied by Wigner and others\cite{wigner}, in which
$V=(1/2)\rmtr\varphi^2$. An important simplifying feature of the analysis in 
\cite{BIPZ} is that $P(\varphi)$ depends only on the eigenvalues of $\varphi$,
and not on the unitary matrix that diagonalizes it. In contrast, the 
probability distribution (\ref{prob}) for non-hermitean matrices depends 
explicitly on the $GL(N)$ matrix $S$ used to diagonalize $\phi=S^{-1} \Lambda 
S$, and $S$ does not decouple. Remarkably however, for the Gaussian $P(\phi)$,
Ginibre \cite{ginibre} managed to integrate over $S$ explicitly and derived an
explicit expression for the probability distribution of the eigenvalues of 
$\phi$. Unfortunately, it is not clear how to integrate over $S$ and derive 
the expression for the eigenvalue probability distribution for non-Gaussian 
distributions of the form (\ref{prob}). In \cite{fz1} this difficulty 
was circumvented by using the method of hermitization.

As an explicit example, the case $V(\pdgp)=2m^2\pdgp + g (\pdgp)^2$ 
was studied in detail in \cite{fz1}. As should perhaps be expected in 
advance, the following behavior in the parameter space $m^2, g>0$ was found: 
for $m^2$ positive, the eigenvalue distribution was disk-like (and 
non-uniform), generalizing Ginibre's work, 
but as $m^2\equiv -\mu^2$ was made more and more negative, a phase transition 
at the critical value \beqast
\mu_c^2 = \sqrt{2g}
\eeqast     
occured, after which the disk fragmented into an annulus. 
The density of eigenvalues was calculated in \cite{fz1} in 
detail.

The paper is organized as follows: In Section 2 we summarize the 
``method of hermitization" \cite{fz}. We present (without derivation) the 
general algorithm for finding the density of eigenvalues associated 
with (\ref{prob}) which was developed in \cite{fz1}, and also add some new
insight into the mechanism behind the ``Single Ring" theorem. We then 
formulate a novel simple criterion on the couplings in $V(\pdgp)$ to 
decide whether the shape of the eigenvalue distribution is a disk or an 
annulus. Finally, we discuss some generic features of the disk-annulus phase 
transition. In particular, we prove that the Green's function associated with 
the hermitean matrix $\pdgp$ (which plays an important role in the 
``hermitization algorithm'' just mentioned) is continuous through the 
disk-annulus phase transition.

In Section 3 we provide simple new expressions for the outer radius $R_{out}$ and for the inner radius $R_{in}$
(in the annular phase) of the eigenvalue distribution of the non-hermitean matrix $\phi$, and for the corresponding boundary values 
$\rho (R_{out})$ and $\rho (R_{in})$ of its eigenvalue density, in terms of the moments 
$$ <\si^k> = \int d\si \si^k \tilde\rho(\si)\quad\quad (k=0, \pm 1, \cdots )$$
of the eigenvalue distribution $\tilde\rho(\si)$ of the hermitean matrix $\pdgp$. Thus, we find that 
$$R_{out}^2 = <\si> $$
and 
$$\rho (R_{out})  = {2R_{out}^2\over <\si^2> - <\si>^2}\,.$$
We see that $R_{out}^2$ is simply the average of $\si$, and the density $\rho(R_{out})$ is inversely proportional to the 
variance of $\si$.

Similarly, we find that in the annular phase, 
$${1\over R^2_{in}} = \langle {1\over \sigma}\rangle $$
and 
$$\rho (R_{in})  = {2R_{in}^{-6}\over <\si^{-2}> - <\si^{-1}>^2}\,.$$
Thus, $R^{-2}_{in}$ is simply the $\si^{-1}$ moment of $\tilde\rho(\si)$, and the density $\rho(R_{out})$ is 
inversely proportional to the variance of $\si^{-1}$.

In Section 4 we verify that the explicit analytic expressions in 
\cite{fz1} concerning the quartic ensemble 
$V(\pdgp)=2m^2\pdgp + g(\pdgp)^2$ are 
consistent with the results of Section 3. We also compare these analytic 
predictions with results of Monte-Carlo simulations of the quartic ensemble 
for various values of $m^2$ and $g$. The numerical results we obtained for 
the eigenvalue distribution in the disk phase and in the 
annular phase, as well as some quantitative features of the disk-annulus 
transition are in good agreement with the analytic predictions in 
\cite{fz1}.

The ``Single Ring Theorem" may seem surprising at first 
sight. Our explanation of why the single ring theorem is not that surprising  
rests upon the simple argument that fragmentation of the 
eigenvalue distribution of $\pdgp$  into several disjoint segments does not 
necessarily imply that the eigenvalues of $\phi$ trace out annuli 
obtained, loosely speaking, by revolving the segments of the eigenvalue 
distribution of $\pdgp$ into the complex plane (see the discussion in 
Section 2). In Section 5 we carry a numerical check of the ``Single Ring" 
theorem for the sextic potential $V(\pdgp)=m^2\pdgp + {\lambda\over 2} (\pdgp)^2 + {g\over 3} (\pdgp)^3 $,
which is the potential of lowest degree for which the eigenvalues of $\pdgp$ may split into more than a single 
segment (in this case, two segments at the most). We generated numerically an ensemble in which the spectrum of 
$\pdgp$ is split into two separated segments, yet we found that the spectrum of $\phi$ is a disk, and not a configuration
of a disk encircled by a concentric annulus, as one would perhaps naively expect by rotating the two-segment spectrum
of $\pdgp$ in the complex plane.

In the Appendix we briefly review the multi-cut phase structure of matrix 
ensembles with generic $V(\pdgp)$, and then specialize to the phase 
structure of the sextic potential ensemble.

\pagebreak

\section{The Method of Hermitization and Non-Gaussian Ensembles}
\setcounter{equation}{0}
Here we very briefly summarize the ``method of 
hermitization" \cite{fz, fz1} in the form of an algorithm, followed by a 
general discussion of the phase structure of the eigenvalue
distribution.

Let us first introduce some notations and definitions. The averaged 
density of eigenvalues 
\beq\label{rho1}
\rho (x,y) =\langle {1\over N} \sum_i \delta(x-\Re \lambda_i)~ \delta (y- \Im
\lambda_i)\rangle
\eeq
of the non-hermitean matrix $\phi$, may be determined from the 
the Green's function associated with $\phi$, namely
\beq\label{greens}
G(z,z^*)=\langle {1\over N} \rmtr {1\over z-\phi}\rangle =
\int d^2 x'~{\rho(x', y')\over z-z'}\,,
\eeq
in terms of which\footnote{We use the following notational conventions:
for $z=x+iy$ we define $\pa\equiv {\partial\over \partial z} = {1\over 2}
\left({\partial\over \partial x} -i{\partial\over \partial y}\right)$
so that $\partial z=1$. Similarly, we define
$\pa^{*}\equiv {\pa\over \pa z^*} = {1\over 2} \left({\pa\over\pa x} +
i{\pa\over\pa y}\right)$, so that $\pa^{*}z^*=1$ and also
$\pa^*(1/z)=\pi\delta^{(2)}(z)$. Finally, we denote $|z|=r$.}
\beq\label{rho11}
\rho(x,y) = {1\over\pi} \pa^{*}~G(z, z^*)\,.
\eeq

The probability distributions (\ref{prob}) studied in this paper are invariant
under $\phi\rightarrow e^{i\alpha}\phi$, rendering 
\beq\label{radrho}
\rho(x,y)\equiv\rho(r)/2\pi
\eeq
circularly invariant. Rotational invariance thus leads to a simpler form of 
the defining formula (\ref{greens}) for $G(z, z^*)$ which reads
\beq\label{circular}
\gam(r) \equiv zG(z,z^*)=\int\limits_0^r r'd r'~\rho(r')\,,
\eeq
whence
\beq\label{rhocirc}
\rho(r) = {1\over r} {d\gam\over dr}\,.
\eeq
Clearly, the quantity $\gam(r)$, which can be thought of as the integrated eigenvalue density, 
is a positive monotonically increasing function,
which satisfies the obvious ``sum-rules" 
\beq\label{sumrules}
\gam(0)=0 \quad\quad {\rm and}\quad\quad \gam(\infty)=1\,.
\eeq
In particular, observe that the first condition in (\ref{sumrules}) insures
that no $\delta(x)\delta(y)$ spike arises in $\rho(x,y)$ when calculating it 
from (\ref{rho11}) with $G(z, z^*)$ given by (\ref{circular}), as it should be.

It was shown in \cite{fz1} that by applying a simple trick, the desired
Green's function of a non-hermitean random matrix $\phi$ could be 
obtained with a minimal amount of work, by
judiciously exploiting the existing literature on random hermitean matrices.
The algorithm, according to \cite{fz1}, for finding the Green's function and 
the averaged eigenvalue density of a non-hermitean random matrix $\phi$ 
drawn from a non-Gaussian ensemble $P(\phi) = (1/Z) e^{-N\rmtr 
V(\phi^\dgg\phi)}$ (Eq. (\ref{prob})) is as follows:

Start with the Green's function\footnote{Of course, $F(w)$ is already 
known in the literature on chiral and rectangular block random hermitean
matrices for the Gaussian distribution\cite{rectangles, bhz, 
bhznpb, rectangles1}, as well as for non-Gaussian probability 
distributions of the form (\ref{prob}) with an arbitrary polynomial 
potential $V(\pdgp)$\cite{cicuta, ambjorn, periwal}.}
\beq\label{FF}
F(w) = \langle {1\over N} \trn {1\over w -\pdgp}\rangle \equiv \int
\limits_0^\infty {\tilde\rho(\si) d\si\over w-\si} \,,
\eeq
where 
\beq\label{dos}
\tilde\rho(\mu) = {1\over N} \langle \trn \delta (\mu -\pdgp)\rangle
\eeq
is the averaged eigenvalue density of $\pdgp$. 
Then, the desired equation for $\gam (r)\equiv zG(z,z^*)$ is 
\beq\label{final}
\gam\left[r^2~F\left({\gam~r^2\over \gam -1}\right) -\gam +1\right] = 0\,.
\eeq
Thus, given $F$ one can solve for $\gam(r)$ using this master equation.

Eq.(\ref{final}) is an algebraic equation for $\gam(r)$ and thus may have 
several $r$ dependent solutions. In constructing the 
actual $\gam(r)$ one may have to match these solutions smoothly into a 
single function which increases monotonically from $\gam(0)=0$ to 
$\gam(\infty)=1$. An explicit non-trivial example of such a procedure is 
the construction of $\gam(r)$ in the disk phase of the quartic 
ensemble\cite{fz1}.

A remarkable property of (\ref{final}) is that it has only two
$r$-independent solutions: $\gam=0$ and $\gam =1$ \cite{fz1}.
Since the actual $\gam(r)$ increases monotonically from 
$\gam(0)=0$ to $\gam(\infty)=1$, we immediately conclude from 
this observation that there can be no more than a single void 
in the eigenvalue distribution. Thus, in the class of models  governed by $P(\phi) = {1\over Z} e^{-N\rmtr V(\phi^\dgg\phi)}$
(Eq. (\ref{prob})), the shape of the eigenvalue 
distribution is either a disk or an annulus, whatever 
polynomial the potential $V(\pdgp)$ is. This result is the ``Single Ring Theorem" of \cite{fz1}.

The ``Single Ring Theorem" may appear counter-intuitive at 
first sight. Indeed, consider a potential $V(\pdgp)$ with 
several wells or minima. For deep enough wells, we expect the 
eigenvalues of $\pdgp$ to ``fall into 
the wells". Thus, one might suppose that the eigenvalue 
distribution of $\phi$ to be bounded by a set of concentric 
circles of radii $0\leq r_1 < r_2 < \cdots < r_{n_{\rm max}}$, 
separating annular regions on which $\rho(r)>0$ from voids 
(annuli in which $\rho(r)=0$.) A priori, it is natural to 
assume that the maximal number of such circular boundaries 
should grow with the degree of $V$, because $V$ may then have 
many deep minima. Remarkably, however, according to the 
``Single Ring Theorem" the number of these 
boundaries is two at the most.

To reconcile this conclusion 
with the a priori expectation just mentioned, note that while the eigenvalues 
of the hermitean matrix $\pdgp$ may split into several disjoint segments 
along the positive real axis, this does not necessarily constrain the 
eigenvalues of $\phi$ itself to condense into annuli. Indeed, the hermitean 
matrix $\pdgp$ can always be diagonalized $\pdgp=U^\dgg\La^2 U$ by a unitary 
matrix $U$, with $\La^2={\rm diag}(\la_1^2, \la_2^2,\cdots,\la_N^2),$ where 
the $\la_i$ are all real. This implies that $\phi=V^\dgg\La U$, with $V$ a 
unitary matrix as well. Thus, the complex eigenvalues of $\phi$ are given by 
the roots of ${\rm det} (z-\La W)=0$, with $W=UV^\dgg$. Evidently, as $W$ 
ranges over $U(N)$ (which is what we expect to happen in the generic case), 
the eigenvalues of $\La W$ could be smeared (in the sense that they would 
not span narrow annuli around the circles $|z|=|\la_i|$.) 

The last argument in favor of the ``Single Ring Theorem" 
clearly breaks down when $W$ fails to range over $U(N)$, which occurs 
when the unitary matrices $U$ and $V$ are correlated. For example, $\phi$   
may be such that $W=UV^\dgg$ is block diagonal, with the upper diagonal block 
being a $K\times K$ unitary diagonal matrix ${\rm diag}(e^{i\om_1}, 
\cdots , e^{i\om_K})$ (and with $K$ a finite fraction of $N$). In the extreme 
case $K=N$, in which $W$ is completely diagonal, 
$W\equiv e^{i\om} = {\rm diag}(e^{i\om_1}, \cdots , e^{i\om_N})$, we see that 
$\phi = U^\dgg e^{i\om}\Lambda U$ is a {\em normal} matrix,\footnote{That is,
$[\phi, \phi^\dgg]=0$.} with eigenvalues ${\rm diag}(e^{i\om_1}\la_1, \cdots ,
e^{i\om_N}\la_N ).$ Thus, normal matrices, or partially normal matrices 
(i.e., the case $K<N$), evade the ``Single Ring'' theorem: if the 
first $K$ eigenvalues $\la_1^2, \la_2^2,\cdots,\la_K^2$ of 
$\pdgp$ split into several disjoint segments along the positive real axis, 
the corresponding eigenvalues of $\phi$ will split into concentric annuli in 
the complex plane obtained by revolving those $\la$-segments. Normal, 
or partially normal matrices are, of course, extremely rare in the 
ensembles of non-hermitean matrices studied in this paper, and do not 
affect the ``Single Ring'' behavior of the bulk of matrices in the ensemble.

We end this section by showing how simple features of $F(w)$ indicate 
whether the domain of the eigenvalue distribution is a disk or an 
annulus.
As is well known \cite{rectangles, ambjorn, periwal}, for $V$ a 
polynomial of degree $p$, the Green's function
$F(w)$ is given by\footnote{Here we assume for simplicity that the
eigenvalues of $\pdgp$ condense into a single segment $[a,b]$. Discussion
of condensation of $\pdgp$ eigenvalues into more segments appears in
the Appendix.} 
\beq\label{FFF} F(w)={1\over 2}V'(w)-P(w)\sqrt{(w-a)(w-b)}\,,
\eeq 
where 
\beq\label{P} P(w) = \sum_{k=-1}^{p-2} c_k~w^k\,. \eeq 
The real 
constants $0 \leq a < b$ and $c_k$ are then determined completely by the
requirement that $F(w)\rightarrow {1\over w}$ as $w$ tends to infinity,
and by the condition that $F(w)$ has at most an integrable singularity as
$w\rightarrow 0$. Thus, if $a>0$, inevitably $c_{-1} = 0$. However, if
$a=0$, then $c_{-1}$ will be determined by the asymptotic behavior for $w$ large. 

According to the ``Single Ring" theorem \cite{fz1}, the 
eigenvalue distribution of $\phi$ is either a disk or an annulus. 
The behavior of $F(w)$ as $w\sim 0$ turns out to be an indicator 
as to which phase of the two the system is in, as we now show:

{\em A. Disk Phase:}~~~In the disk phase we expect that $\rho(0)>0$, as in Ginibre's case. 
Thus, from (\ref{rhocirc}) $\rho(r) = (1/r) (d\gam/dr) \equiv 2 
(d\gam/dr^2)$ and from the first sum rule $\gam(0)=0$ in (\ref{sumrules}) we 
conclude that 
\beq\label{near0}
\gam(r)\sim {1\over 2} \rho(0) r^2
\eeq
near $r=0$. 
Therefore, for $r$ small, 
(\ref{final}) yields 
\beq\label{near00}
F\left(-{\rho(0) r^4\over 2} + \cdots \right)\sim - {1\over r^2}\,,
\eeq 
namely, $F(w)\sim 1/\sqrt{w}$ for $w\sim 0$, as we could
have anticipated from Ginibre's case.~\footnote{In the Gaussian case, 
$V=\pdgp$, we have $2\sqrt{w}F(w) = \sqrt{w} - \sqrt{w -4}$, whence the roots 
of (\ref{final}) are $\gam=0, 1$ and $r^2$. We note that $\gam=0$ is 
unphysical, $\gam=r^2$ ({\em i.e.,} $G=z^*$) corresponds to Ginibre's 
disk\cite{ginibre}, and $\gam=1$ is the solution outside the disk.} This means
that in the disk phase we must set $a=0$ in (\ref{FFF}). Consequently, in the 
disk phase $c_{-1}$ does not vanish. We can do even better: paying 
attention to the coefficients in (\ref{FFF}) and (\ref{P}) (with $a=0$) we 
immediately obtain from (\ref{near00}) that 
\beq\label{rho0}
c_{-1} = \sqrt{\rho(0)\over 2 b}\,.
\eeq

{\em B. Annular Phase:} In the annular phase $\gam(r)$ must clearly  
vanish identically in the inner void of the annulus.  Thus, 
(\ref{final}) implies that $F(w)$ cannot have a pole at $w=0$, and therefore from 
(\ref{FFF}) we must have $c_{-1}\sqrt{ab}=0$. Thus, the annulus must 
arise for $c_{-1} = 0$ (the other possible solution $a=0, c_{-1}\neq
0$ leads to a disk configuration with $\gam=0$ only at $r=0$, as we just 
discussed.)

Thus, to summarize, in the disk phase $F$ has the form 
\beq\label{Fdisk}
F_{disk}(w)={1\over 2}V'(w)-\left(\sqrt{{\rho(0)\over 2b}}\,w^{-1} + c_0 + c_1 ~w 
+\cdots + c_{p-2}~w^{p-2}\right)\sqrt{w(w-b)}\,,
\eeq
while in the annular phase it has the form
\beq\label{Fannulus}
F_{annulus}(w)={1\over 2}V'(w)-\left(c_0 + c_1 ~w +\cdots 
+c_{p-2}~w^{p-2}\right)\sqrt{(w-a)(w-b)}\,. \eeq
Having determined $F(w)$ in this way, i.e., having determined the various unknown parameters 
in (\ref{Fdisk}) or in (\ref{Fannulus}), we substitute it into (\ref{final}) and 
find $G(z, z^*)$. We can thus calculate the density of eigenvalues $\rho(r)$ explicitly 
for an arbitrary $V$. 

We now turn to the disk-annulus phase transition. An important feature of this transition is that $F(w)$ is continuous
through it. To see this we argue as follows: By tuning the couplings in $V$, we can induce a phase transition from the 
disk phase into the annular phase,
or vice versa. Note, of course, that we can parametrize any point in the disk phase either by the set of couplings in
$V$ or by the set of parameters $\{c_{-1}, c_0,  \cdots c_{p-2}; b\}$ in (\ref{Fdisk}). The ``coordinate transformation" between
these two sets of parameters is encoded in the asymptotic behavior of $F(w)$. Similarly, 
we can parametrize any point in the annular phase either by the set of couplings in
$V$ or by the set of parameters $\{c_0, c_1 \cdots c_{p-2}; a, b\}$ in (\ref{Fannulus}). 
Due to the one-to-one relation (in a given phase, once we have established it is the stable one) between the couplings
in $V$ and the parameters in $F(w) - {1\over 2} V'(w)$ (namely, the $c_n$'s and the locations of the branch points of $F(w)$), we can 
describe the disk-annulus transition in terms of the latter parameters (instead of the couplings in $V$). Clearly, the 
transition point is reached from the disk phase when $\rho(0) = 0$, that is, when $c_{-1}$ in (\ref{Fdisk}) vanishes:
\beq\label{cminus1crit}
c_{-1}^{crit} = 0\,.
\eeq
Similarly, the transition point is reached from the annular phase when the lower branch point $a$ in (\ref{Fannulus}) vanishes. 
Thus, e.g., in a transition from the disk phase into the annular phase, $F_{disk}(w)$ in (\ref{Fdisk}) would cross-over 
continuously into $F_{annulus}(w)$ in (\ref{Fannulus}) through a critical form 
\beq\label{Fcrit}
F_{crit}(w)={1\over 2}V_{crit}'(w)-\left(c^{crit}_0 + c^{crit}_1 ~w +\cdots 
+c^{crit}_{p-2}~w^{p-2}\right)\sqrt{w(w-b^{crit})}\,.\eeq

The continuity of $F(w)$ through the transition was demonstrated explicitly 
in \cite{fz1} for the quartic ensemble $V(\pdgp) = 2m^2 \pdgp + g(\pdgp)^2$ 
(see also Section 4). 

This discussion obviously generalizes to cases when $F(w)$
has multiple cuts, which correspond to condensation of the
eigenvalues of $\pdgp$ into many segments. If $w=0$ is a branch
point of $F(w)$, that is, if the lowest cut extends to the
origin, we are in the disk phase,
\beq\label{Fdisk1}
F_{disk}(w)={1\over 2}V'(w)-\left(c_{-1}w^{-1} + c_0 + c_1 ~w 
+\cdots + c_{p-2}~w^{p-2}\right)\sqrt{w(w-b_1)\cdots (w-b_n)}\,,
\eeq
with $0< b_1 < \cdots < b_n $. The relation (\ref{rho0})
then generalizes to 
\beq\label{rho00}
c_{-1} = \sqrt{{\rho(0) \over 2 (-1)^{n+1} \prod_{k =1}^n b_k}}\,.
\eeq
Since $c_{-1}$ must be real we conclude that such a configuration
exists only for $n$ odd. 

If the lowest branch 
point in $F(w)$ is positive, we are in the annular phase with 
\beq\label{Fannulus1}
F_{annulus}(w)={1\over 2}V'(w)-\left(c_0 + c_1 ~w +\cdots 
+c_{p-2}~w^{p-2}\right)\sqrt{(w-a)(w-b_1)\cdots (w-b_n)}\,. \eeq
The phase transition would occur when the couplings in $V(\pdgp)$ are tuned 
such that $F_{disk}(w)$ and $F_{annulus}(w)$ match continuously, as 
was described in the previous paragraph.

\pagebreak
\section{Boundaries and Boundary Values}
\setcounter{equation}{0}
Remarkably, with a minimal amount of effort, and based on the mere
definition of $F(w)$ (Eq. (\ref{FF}), which we repeat here for convenience)
\beq\label{FFrep}
F(w) = \langle {1\over N} \trn {1\over w -\pdgp}\rangle \equiv \int
\limits_0^\infty {\tilde\rho(\si) d\si\over w-\si} \,,
\eeq     
we are able to derive simple expressions for the location of the 
boundaries of the eigenvalue distribution and also for the 
boundary values of $\rho(r)$ in terms of the moments of $\tilde\rho(\si)$, which, we remind the reader, 
is the density of eigenvalues for a hermitean matrix problem. 

To this end it is useful to rewrite our master formula (\ref{final}) for
$\gam (r)$ as 
\beq\label{fin}
w F(w) = \gam
\eeq
with 
\beq\label{w}
w = {\gam r^2\over \gam -1}\,.
\eeq

We start with the outer edge $r=R_{out}$ (either in the disk phase or
in the annular phase.) Near the outer edge $\gam\rightarrow 1-$, and thus 
$w\rightarrow -\infty$. We therefore expand $F(w)$ in powers of $1/w$ and 
obtain from (\ref{FFrep})-(\ref{w})
\beq\label{finsum1}
{<\si>\over r^2} + {\gam -1\over \gam r^4 } <\si^2> + {(\gam -1)^2 \over 
\gam^2 r^6 } <\si^3> + \cdots  = \gam\,, \eeq  
where
\beq\label{moments}
<\si^k> = \int\limits_0^\infty \tilde\rho(\si) \si^k d\si
\eeq
are the moments of $\tilde\rho(\si)$ (which is of course normalized to 1.)
For the class of models we are interested in here, all the 
moments $<\si^k>,~ k\geq 0$ are clearly finite.\footnote{$\tilde\rho(\si)
\equiv (1/\pi){\rm Im} F(\si -i\epsilon)$ is supported along a finite segment 
(or segments), and its singularity at $\si=0$ is no worse than 
$\si^{-1/2}$.} Thus, at the outer edge $r=R_{out}$ (where of 
course $\gam(R_{out}) =1$), all terms with $<\si^k>,\,  k\geq 2$ drop out of 
(\ref{finsum1}) and we obtain
\beq\label{rout}
R_{out}^2 = <\si> \,.
\eeq
Namely, $R_{out}^2$ is simply the first moment of $\tilde\rho(\si)$.

We now calculate the boundary value $\rho(R_{out})$. Approaching $R_{out}$
from the inside, we substitute $\gam = 1-f$ and $r^2 = R_{out}^2 (1-\delta)$
(with $f, \delta <<1$) in (\ref{finsum1}). After some work we obtain
$f = {<\si>^2 \over <\si^2> -<\si>^2} \delta + \cO (\delta^2)$, namely,
\beq\label{gamf}
\gam = 1 - {<\si>^2 \over <\si^2> -<\si>^2}\,\, \delta + \cO (\delta^2)\,.
\eeq
Thus, from $\rho(r) = 2 (d\gam/dr^2)$ (Eq. (\ref{rhocirc})) and 
(\ref{rout}) we find 
\beq\label{rhoout}
\rho (R_{out})  = {2R_{out}^2\over <\si^2> - <\si>^2}\,.
\eeq
The density $\rho(R_{out})$ is inversely proportional to the variance
of $\si$! 

For the $\tilde\rho(\si)$ under consideration here, $<\si^2>$, 
and consequently $\rho(R_{out})$, are always finite. Outside the
boundary $\rho(r)$ vanishes identically, of course, and thus, $\rho(r)$ 
always ``falls off a cliff" at the boundary, for all probability 
distributions of the form (\ref{prob}) with $V$ polynomial. 
It would be thus interesting to study circularly invariant matrix
ensembles $P(\pdgp)$ such that the eigenvalue distribution 
$\tilde\rho(\si)$ of $\pdgp$ has a finite $<\si>$ but an infinite 
$<\si^2>$. Then $\rho(R_{out})$ would vanish. This would naturally raise 
the question whether in such situations, $\rho(r)$ behaves universally 
near the edge (that is, if near the edge it vanishes like $(R_{out} - 
r)^\epsilon$ with $\epsilon$ being some universal exponent). 
We do not pursue this question further in this paper.

We now turn to the annular phase, and focus on the inner edge 
$r=R_{in}$ of the annulus. According to the discussion at the end of 
Section 2 (see Eq. (\ref{Fannulus}) and the discussion above it), $a>0$ in 
(\ref{FFF}), and thus $F(w)$ is analytic in the domain $|w|<a$. 
Expanding (\ref{FFrep}) in powers of $w$, we obtain from 
(\ref{fin}) 
\beq\label{fin2}
{1-\gam\over r^2} - \langle {1\over\si}\rangle = 
w\langle {1\over \si^2}\rangle +
w^2\langle {1\over \si^3}\rangle + \cdots\,.
\eeq
A little above the inner radius, into the annulus, 
clearly $\gam \rightarrow 0$ and $w\rightarrow 0-$ in (\ref{w}). 
Thus, setting $w=0$ in (\ref{fin2}) we obtain
\beq\label{rin}
{1\over R^2_{in}} = \langle {1\over \sigma}\rangle\,.
\eeq
$R^{-2}_{in}$ is simply the $\si^{-1}$ moment of $\tilde\rho(\si)$.

We can now calculate the boundary value $\rho (R_{in})$. Near the inner
edge we parametrize $r^2= R^2_{in} (1+\delta)$ with $\delta <<1$ (and of 
course, $\gam << 1$ to begin with.) Since $\tilde\rho(\si)$ obviously 
vanishes 
for $\si<a$, all moments $<\si^{-k}>$ in (\ref{fin2}) are finite. Thus, 
dropping all terms with $<\si^{-k}>, k\geq 3$ in (\ref{fin2}), we obtain
after some work
\beq\label{gamin}
\gam = R^{-4}_{in} {\delta\over <\si^{-2}> - <\si^{-1}>^2} + \cO 
(\delta^2)\,. \eeq
It then follows from (\ref{rhocirc}) and (\ref{rin}) that 
\beq\label{rhoin}
\rho (R_{in})  = {2R_{in}^{-6}\over <\si^{-2}> - <\si^{-1}>^2}\,.
\eeq
The density $\rho(R_{out})$ is inversely proportional to the variance
of $\si^{-1}\,.$

From (\ref{FFF}) (or (\ref{Fannulus}) ) we learn that in the annular 
phase $\tilde\rho_{annulus}(\si) \equiv {1\over \pi} {\rm Im} F(\si-i\epsilon)
= {\rm polynomial}(\si)\sqrt{(\si - a)(b - \si)}\,,~0<a<\si<b$ (and 
vanishes elsewhere.) Thus, $<1/\si^k> \equiv \int\limits_{a}^{b}
(\tilde\rho(\si)/\si^k)d\si\,,~k=1,2$ are finite. Therefore, 
$\rho_{annulus}(\si)$ jumps from zero (in the inner void of the annulus)
to a finite value at the inner edge $R_{in}$. Note, however, 
that when $a\rightarrow 0$, that is, in the annular to disk transition,
$<1/\si>$ remains finite, but $<1/\si^2>$ diverges like $1/\sqrt{a}$. (For a 
particular example see Eq. (\ref{momentsannQ}).) 
Thus, from (\ref{rin}) we see that $R_{innner} (a=0)$, the {\em critical} 
inner 
radius, is finite. The annulus starts up with a finite inner radius. Also,
in this limit, we see from (\ref{rhoin}) that $\rho(R_{in})$ vanishes 
like $\sqrt{a}$. As we approach the annulus-disk transition, the 
discontinuity in $\rho(r)$ at the (finite) inner edge disappears.

We saw at the end of Section 2 (see Eqs. (\ref{Fdisk})-(\ref{Fcrit})) that $F(w)$ is continuous through the disk-annulus phase 
transition. Thus, our master formula $wF(w)=\gam $ to determine $\gam(r)$ (Eq. (\ref{fin})) is also continuous through the 
transition. Consequently, $\rho(r) = (1/r) (d\gam/dr)$ must remain continuous through the disk-annulus transition, and
has (at the transition) the universal behavior described in the previous paragraph.

\pagebreak
\section{Phase Transitions in the Quartic Ensemble}
\setcounter{equation}{0}

The disk-annulus transition in the quartic ensemble 
\beq\label{quartic}
V(\pdgp) = 2 m^2\pdgp + g (\pdgp)^2
\eeq
was studied in detail in \cite{fz1}. The annular eigenvalue 
distribution $\rho_{annular}(r)$ and the disk eigenvalue distribution 
$\rho_{disk}(r)$ for this ensemble were calculated explicitly in \cite{fz1}.
According to the expressions given in \cite{fz1}, as the critical point is 
approached from the annular phase, $\rho_{annular}(r)$ behaves 
precisely as described in the paragraph following Eq. (\ref{rhoin}) at the 
end of the previous section (see also Eq.(\ref{rhoann}) below, at $\mu=\mu_c$.) 
Also according to \cite{fz1}, as the critical point is approached from the 
disk phase, $\rho_{disk}(r)$ gets completely depleted inside a region of 
radius $R_{in}(\mu_c)$ (remaining continuous at $r=R_{in}(\mu_c)$. 
See Eq. (\ref{rhodiskcrit}) below.) Thus, $\rho(r)$ for the quartic ensemble 
is continuous through the disk-annulus transition. 

In this section we verify the expressions (\ref{rout}), (\ref{rhoout}), 
(\ref{rin}) and (\ref{rhoin}) for $R_{out}$, $\rho (R_{out})$, $R_{in}$ and 
$\rho (R_{in})$ for the quartic ensemble (\ref{quartic}) against the 
explicit expressions for these quantities given in \cite{fz1}, and also 
provide ample numerical results concerning the disk phase, the annular 
phase, and the transition between them, in support of the analytical results.
In what follows we have omitted many technical details that can be found in 
\cite{fz1}.

\subsection{The Disk Phase}

For $m^2>-\sqrt{2g}$ the density of eigenvalues is a disk.
According to \cite{fz1} we have
\beq\label{FQ}
F(w)=m^2 + gw - \left({c\over w} + g\right)~\sqrt{w (w-b)}
\eeq
with 
\beq\label{parameters1}
c = {2m^2 +\sqrt{m^4 + 6g}\over 3}, \quad 
{\rm  and}\quad b={-2m^2 + 2\sqrt{m^4 + 6g}\over 3g}\,.
\eeq

According to Eqs. (5.8) and (5.9) in \cite{fz1}, the eigenvalue density 
in this phase is 
\beq\label{rhodisk}
\rho_{disk}(r) = 2m^2 + 4gr^2 + 2[{\rm sgn}~({b\over 4}-r^2)] ~
{bc^2-(m^2+2gr^2)[1+2(m^2r^2+gr^4)]\over\sqrt{[1+2(m^2r^2+gr^4)]^2 - 4bc^2r^2}}
\eeq
inside a disk of radius $R_{out}$, where 
\beq\label{routdisk}
R_{out}^2 = {bc^2 - 2m^2\over 2g} = {(m^4+6g)^{3/2}-m^2(m^4 + 9g)\over 27 g^2}\,.
\eeq
Thus, from (\ref{rhodisk}) and (\ref{routdisk}) we have \footnote{It is 
straight forward to show that ${\rm sgn}~({b\over 4}-R_{out}^2)=-1$ in
(\ref{rhodisk}): just substitute $m^4 = A^2 {\rm cos}^2 \alpha$ and
$6g = A^2 {\rm sin}^2 \alpha$ and obtain that $R_{out}^2 - {b\over 4} = 
(A^3/108g^2) (1-{\rm cos}\alpha )^3$.} 
\beq\label{rhoroutdisk}
\rho_{disk}(R_{out}) = {4g(bc^2 - 2m^2)\over 2g - bc^2(bc^2 - 2m^2)} = 
{4gR_{out}^2\over 1-2R_{out}^2 ( g R_{out}^2 + m^2)}\,.
\eeq

These results should be compared with the predictions of Section 3. 
From (\ref{FQ}) we can read-off the density of eigenvalues 
$\tilde\rho(\sigma) = (1/\pi){\rm Im} F(\sigma - i\epsilon)$ of $\pdgp$ as
\beq\label{dosQ}
\tilde\rho(\sigma ) = {1\over \pi} \left({c\over\sigma} + g\right)\sqrt{\sigma
(b-\sigma)}\eeq
for $0\leq\sigma\leq b$, and zero elsewhere. We can readily check that 
(\ref{dosQ}) is properly normalized to 1. 

The first two moments of (\ref{dosQ}) are 
\beq\label{1momentdiskQ}
<\si> = {1\over 2}\left({b\over 2}\right)^2 \left( c + {gb\over 2}\right) = 
{(m^4+6g)^{3/2}-m^2(m^4 + 9g)\over 27 g^2}\,,
\eeq
and
\beq\label{2momentdiskQ}
<\si^2> = {1\over 8}\left({b\over 2}\right)^4 \left( {8c\over b} + 5g\right) =
{27g^2 + 18gm^4 + 2m^8 -2m^2 (6g + m^4)^{3/2}\over 54g^3} \,.
\eeq

Thus, 
\beqra\label{variance}
&&<\si^2> - <\si >^2 = -{b^3\over 256}\left[4 b c^2  + bg (b^2 g - 10) + 4 c ( b^2 g -4)\right]\nonumber\\{}\nonumber\\
 &=& {297g^3 + 108 g^2 m^4 -18 g m^8 -4m^{12} -2m^2(9g -2m^4)(6g + m^4)^{3/2}\over 1458 g^4}\,.
\eeqra

Comparing (\ref{routdisk}) and (\ref{1momentdiskQ}) we immediately verify 
(\ref{rout}), $R^2_{out} = <\si>$. After some additional work, using 
(\ref{variance}) and (\ref{routdisk}) in (\ref{rhoout}), we can see that 
(\ref{rhoout}), namely, that $\rho (R_{out}) = 2 R^2_{out}/(<\si^2>-<\si>^2)$, coincides with (\ref{rhoroutdisk}).

\subsubsection{Numerical Results for the Disk Dhase}

We have generated numerically random matrix ensembles corresponding to the 
quartic potential (\ref{quartic}) in the disk phase, for $m^2=1$ fixed and 
for various values of the coupling $g$ (and for various sizes of matrices),
and measured $\rho_{disk}(r)$ for these realizations.

The generation of the matrices was done by a standard Metropolis 
Monte Carlo approach.  A random change had been suggested in the real and 
imaginary parts of one of the elements of $\phi$ and then the change in 
$V(\phi)$ was evaluated.  This ``move'' was accepted unconditionally if 
$V$ were decreased, and with probability $p=e^{-\Delta V}$ if $V$ were 
increased.  
General theorems on Monte Carlo then guaranteed that the resulting 
probability distribution of $\phi$ was the desired one.  After the 
matrices were generated, their eigenvalues were determined with a standard 
solver from the LAPACK library.  We tuned the size of the suggested
changes in $\phi$ so that the acceptance rate was about one--half, and
monitored the equilibration and autocorrelation times to ensure our
starting configuration had evolved properly and error bars have were 
accurate. In particular, the local changes in $\phi$ made the matrices 
correlated over some number of random changes, however, local changes 
also allowed one to employ various tricks to evaluate the change in $V$ 
rapidly.

In Figure 1 we display our numerical results for $\rho_{disk} (r)$ for 
128x128 dimensional matrices, and compare them to the analytical large-$N$ 
result (\ref{rhodisk}) of \cite{fz1}. (As a trivial check of our numerical 
code, we also included in this figure the results for the gaussian (Ginibre) 
ensemble.)

Evidently, the agreement between the numerical and the analytical 
results is good. Note the finite-$N$ effects near the edge of the disk.

\subsection{The Annular Phase}
For $m^2 <-\sqrt{2g}$, the stable eigenvalue distribution is annular. 
For convenience, let us switch notations according to $m^2=-\mu^2$, and also 
write $\mu_c^2=\sqrt{2g}$.

According to \cite{fz1} we have 
\beq\label{FANN}
F(w)=m^2 + gw - g~\sqrt{(w-a) (w-b)}
\eeq
with 
\beq\label{parameters2}
a={\mu^2\over g} -\sqrt{{2\over g}}, \quad {\rm  and}\quad 
b={\mu^2\over g} +\sqrt{{2\over g}}\,.
\eeq
We see that $a= (2/\mu_c^4)(\mu^2-\mu_c^2)$ which is positive for 
$\mu^2>\mu_c^2$, as it should be, by definition.

According to Eqs. (5.16) - (5.19) in \cite{fz1}, the eigenvalue density in 
this phase is 
\beq\label{rhoann}
\rho_{annulus} (r) = 8g\left(r^2 -{\mu^2\over 2g}\right)= 
8g\left(r^2 -{\mu^2\over\mu_c^4}\right)
\eeq
inside an annulus $R_{in}\leq r\leq R_{out}$, where
\beq\label{ranin}
R_{in}^2={\mu^2 + \sqrt{\mu^4-2g}\over 2g} = 
{\mu^2 + \sqrt{\mu^4-\mu_c^4}\over \mu^4_c}
\eeq
and 
\beq\label{ranout}
R_{out}^2={\mu^2\over g} = {2\mu^2\over \mu_c^4}\,.
\eeq
Thus, we see immediately that 
\beqra\label{rhoannbound}
\rho (R_{in}) &=& 4\sqrt{\mu^4-\mu_c^4}\nonumber\\{\rm and}\nonumber\\
\rho (R_{out}) &=& 4\mu^2\,.
\eeqra
Note that $\rho (R_{in})=0$ at $\mu=\mu_c$, as expected. Also note that 
the critical annulus has a finite inner radius: 
$R_{in}^2 (\mu_c) = 1/\mu_c^2 > 0 $. 

We now compare these results with the predictions of Section 3. From 
(\ref{FANN}) we read-off the density of eigenvalues of $\pdgp$: 
\beq\label{dosQFANN}
\tilde\rho(\sigma ) = {g\over \pi} \sqrt{(\sigma -a ) (b-\sigma)}
\eeq
for $a\leq\sigma\leq b$, and zero elsewhere. We can check that 
(\ref{dosQFANN}) is properly normalized to 1. 

The relevant moments of (\ref{dosQFANN}) are 
\beqra\label{momentsannQ}
<\si> &=& {g\over 2}\left({a+b\over 2}\right)\left({b-a\over 2}\right)^2 = 
{\mu^2\over g}\nonumber\\{}\nonumber\\
<\si^2> &=& {g\over 2}\left({b-a\over 2}\right)^4 \left[\left({b+a\over b-a}\right)^2 + {1\over 4}\right] = {\mu^4\over g^2} + {1\over 2g}\nonumber\\{}\nonumber\\
<{1\over\si}> &=& {g\over 2}(\sqrt{b}-\sqrt{a})^2 = \mu^2 - \sqrt{\mu^4 - 2g}
\nonumber\\{\rm and }\nonumber\\
<{1\over\si^2}> &=& {g\over 2}{(\sqrt{b}-\sqrt{a})^2\over\sqrt{ab}} = 
g\, {\mu^2 - \sqrt{\mu^4 - 2g}\over \sqrt{\mu^4 - 2g}}\,.
\eeqra

Comparing (\ref{ranin}), (\ref{ranout}) and the first and third equations in 
(\ref{momentsannQ}), we verify (\ref{rout}) and (\ref{rin}) straightforwardly. 

Further, we find from (\ref{momentsannQ}) that 
\beqra\label{variances}
<\si^2> - <\si>^2 &=& {1\over 2g}\nonumber\\{\rm and}\hfill\nonumber\\
<{1\over\si^2}> - <{1\over \si} >^2 &=& g-2\mu^4 +\mu^2 {2\mu^4-3g\over \sqrt{\mu^4-2g}}\,.
\eeqra
Thus, comparing with (\ref{rhoannbound}) we find that 
$$ {2R_{out}^2\over <\si^2> - <\si>^2} = 4\mu^2 = \rho (R_{out})$$ 
and 
$${2R_{in}^{-6}\over <\si^{-2}> - <\si^{-1}>^2} = 4\sqrt{\mu^4-\mu_c^4} = 
\rho (R_{in})\,,$$ 
and verify (\ref{rhoout}) and (\ref{rhoin}) for the annular phase.

\subsubsection{Numerical Results for the Annular Phase}

In Figures 2.a-2.c we display our numerical results for 
$\rho_{annulus} (r)$ for matrices of various sizes, 
and compare them to the analytical large-$N$ result (\ref{rhoann}) of 
\cite{fz1}. In these figures we hold $\mu^2=0.5$ fixed, and increase $g$ 
from $0.025$ to $0.1$\footnote{Here we have $\mu^2= 0.5 = 
\mu_c^2/2\sqrt{2g}$. Thus increasing $g$ as indicated in the text brings us 
closer to the disk-annulus phase transition.}.

\subsection{The Disk-Annulus Phase Transition}

The phase boundary separating the disk phase and the annular phase 
in the $m^2-g$ plane is the curve $m^2 = -\sqrt{2g}$. 

Consider approaching this boundary from within the disk phase by setting 
$m^2=-\sqrt{2g} + \delta$, with $\delta$ positive and small. Then, 
using (\ref{parameters1}), we find to first order in
$\delta$ that $c=\delta/2$ and $b=2\sqrt{(2/g)} - \delta/g$.
In particular, at the phase boundary itself $c=0$, 
in accordance with (\ref{cminus1crit}) and (\ref{Fcrit}). It was shown in \cite{fz1}
that as one approaches the critical point $m^2 = -\sqrt{2g}$ from the disk 
phase, the density of eigenvalues of $\phi$ approches the particularly simple 
critical configuration 

\beq\label{rhodiskcrit}
\rho_{crit} (r) = \left\{\begin{array}{c} 0\quad ,\quad r^2<1/\sqrt{2g}\,\,\\
{}\\
8g (r^2 - {1\over\sqrt{2g}}) \quad ,\quad 1/\sqrt{2g}<r^2<\sqrt{2/g}\,,
\end{array}
\right.
\eeq
Thus, as we decrease 
$\delta$ to zero, $\rho_{disk}(r)$ (Eq. (\ref{rhodisk})) becomes increasingly 
depleted inside the disk $r^2< b(\delta)/4$, reaching complete depletion at 
$\delta=0$, at 
which point the disk breaks into an annulus. We also note that at the phase 
boundary (\ref{FQ}) reads 
\beq\label{FQcrit}
F(w)=-\sqrt{2g} + gw - g~\sqrt{w \left(w-2\sqrt{{2\over g}}\right)}\,.
\eeq

Consider now approaching the phase boundary $m^2 = -\sqrt{2g}$ from within the 
annular phase.  Thus, we set $\mu^2=\sqrt{2g} + \delta$, with $\delta$
positive and small. Then, since all the expressions in (\ref{parameters2}) are
linear in $\mu^2$, we find that $a=\delta/g$ and $b=2\sqrt{(2/g)} + \delta/g$.
In particular, at the phase boundary itself $a=0$, and $b=1/\sqrt{2g}$. 
Therefore, at the phase boundary (\ref{FANN}) reads 
$$ F(w)=-\sqrt{2g} + gw - g~\sqrt{w \left(w-2\sqrt{{2\over g}}\right)}\,,$$
which coincides with (\ref{FQcrit}). Thus, $F(w)$ (and consequently, the 
eigenvalue density of $\phi^\dgg\phi$) is also continuous at the transition,
in accordance with (\ref{Fcrit}).

Note from (\ref{ranin}) and (\ref{ranout}) that at the transition 
$R_{in}^2=1/\mu_c^2=1/\sqrt{2g}$ is {\em finite}, and coincides 
with the radius (squared) of the depleted region in the disk configuration 
(\ref{rhodiskcrit}). Right at the transition, the disk breaks into an annulus with a finite hole! Note also that $R_{out}^2 = 2/\mu_c^2 = 
\sqrt{2/g}$, which coincides with the disk's $R_{out}^2$  
at the phase boundary. Thus, at the phase boundary $\mu^2=\mu_c^2$ 
(\ref{rhoann}) coincides with (\ref{rhodiskcrit}), namely, $\rho(r)$
is continuous at the transition from the disk phase to the annular phase.

\subsubsection{Numerical Simulation of the Disk-Annulus Phase Transition}

We have measured the density of eigenvalues 
$\rho (r)$ of matrices $\phi$ of size 128x128, taken from the 
the quartic ensemble with $\mu^2=0.5$ and for 
$g = 0.025, 0.05, 0.1, 0.125, 0.15$ and $0.175$. The results are 
displayed on Figure 3. 

For these values of $g$, we start in the annular phase 
at the lowest value of $g$. For our set of parameters we have $\mu^2= 0.5 = 
\mu_c^2/2\sqrt{2g}$. Thus, increasing $g$ (while keeping $\mu^2$ fixed at 
$0.5$) brings us closer to the disk-annulus phase transition, which occurs 
(at large $N$) at $g_c=0.125$. Increasing $g$ 
beyond that, puts us into the disk phase.  

The first three profiles on the right in Figure 3 belong to the annular phase. 
Their behavior is consistent with our discussion in Section 4.2 of the 
annular phase. Indeed, as $g$ increases towards the transition point at 
$g_c=0.125$, these three graphs exhibit the expected decrease of 
$R^2_{in} = (\mu^2 + \sqrt{\mu^4-\mu^4_c})/\mu_c^4$ \,(Eq. (\ref{ranin}), 
with $\mu_c^2=\sqrt{2g}$) and the decrease of 
$R_{out}^2 = 2\mu^2/\mu_c^4$ \,(Eq. (\ref{ranout})).

The critical density profile, corresponding to 
$g_c=0.125$, is the fourth profile (from the right). For our choice of 
parameters, the theoretical boundary radii of the critical annulus, i.e., at 
$g=0.125$, are $R_{in}^{crit} = 1/\mu_c = \sqrt{2}$ and 
$R_{out}^{crit}=\sqrt{2}/\mu_c=2$. These boundary values fit nicely with 
the features of the critical profile in Figure 3.

Finally, the last two profiles in Figure 3 have pronounced tails extending 
to $r=0$ and thus belong to the disk phase.

\pagebreak

\section{Phase Transitions in the Sixth Order Potential and the ``Single Ring" Theorem}
\setcounter{equation}{0}

The sextic potential 
\beq\label{sextic}
V(\phi^\dgg\phi) = m^2 \pdgp +{\lambda\over 2} (\pdgp)^2 + {g\over 3} (\pdgp)^3\,
\eeq
is the potential of lowest degree in (\ref{prob}) for which the eigenvalues of $\pdgp$ may split into more then a single 
segment. In fact, it is easy to see that there can be at most two eigenvalue segments in the spectrum of $\pdgp$.

The qualitative features of the support of the eigenvalue density 
associated with (\ref{sextic}) can be deduced by moving the cubic 
$V(x) = m^2 x + {\lambda\over 2} x^2 + {g\over 3} x^3$
around in the plane (i.e., by varying its couplings, fixing, say $g=1$), 
and concentrating on
$x\geq 0$. It is obvious from such considerations that there are three 
qualitatively different phases in the spectrum
of $\pdgp$. In two of the phases the eigenvalues $s_i$ of $\pdgp$ live in a single segment. In one of these single 
segment phases, the segment includes the origin ($0\leq s\leq a$), but in the other it does not ($0<a\leq s \leq b$). 
Following the general discussion in the last part of Section 2, we would 
expect that the spectrum of $\phi$ itself is a disk in the first case, and an 
annulus in the second case.

In the third phase of $\pdgp$, there are two 
segments, one of which hits the origin ($\{0\leq s \leq a\}
\bigcup\{b\leq s \leq c\}).$ (There is no two-segment phase of $\pdgp$ which 
does not include the origin.) Thus, according to the discussion in the 
last part of Section 2, the non-hermitean matrix $\phi$ is expected to be in 
the disk phase in this case (rather than having its eigenvalue fill in a disk 
surrounded by a concentric annulus), in accordance with the ``Single Ring" 
theorem.

In this short section we limit our discussion to the two-segment phase of
$\pdgp$. (A rather detailed sketch of the analytical conditions that 
determine the whole phase structure of the sextic ensemble (\ref{sextic}) is 
given in the Appendix.) Our purpose here is to demonstrate numerically 
the ``Single Ring'' Theorem for the eigenvalue distribution of matrices 
$\phi$ taken from the sextic ensemble (\ref{sextic}).  To this end, we have 
to identify points well within the phase in which the eigenvalues of 
$\pdgp$ split into two segments.

We used the formalism of the Appendix to choose two ensembles in the 
two-segment phase of $\pdgp$, for which we verified that the 
eigenvalues of $\phi$ formed a disk. The results for these ensembles are 
displayed in Figures 4 and 5.

Figure 4 shows the scatter plot of the eigenvalues of $\phi$ together with 
the density of eigenvalues $\tilde\rho(s)$ of $\pdgp$ for (\ref{sextic}) with 
couplings $m^2=7.372, \lambda=-6.116 $ and $g=1.372$. As can be seen on the 
right part of Figure 4, for these couplings, the eigenvalues of 
$\pdgp$ live in two separated segments: 
$\{0\leq s \leq a=1\}\bigcup\{b=2\leq s \leq c=3\}$. The solid line there is 
the large $N$ theoretical curve, which was plotted 
according to the analysis we have described in the Appendix (see the 
discussion following (\ref{2cutdisk})). Evidently, the spectrum of 
$\phi$ is a disk, despite the split support of $\tilde\rho(s)$, in accordance 
with the ``Single Ring'' Theorem.

Figure 5 is similar to Figure 4, but for (\ref{sextic}) with 
couplings $m^2=6.403, \lambda=4.184$ and $g=0.713$, for which the 
eigenvalues of $\pdgp$ live in the segments $\{0\leq s \leq a=1\}\bigcup
\{b=3\leq s \leq c=4\}.$ The spectrum of $\phi$ remains a disk, 
even though the two segments of the support of $\tilde\rho(s)$ are more 
separated than in Figure 4.


\newpage
\setcounter{equation}{0}
\setcounter{section}{0}
\renewcommand{\theequation}{A.\arabic{equation}}
\renewcommand{\thesection}{Appendix A}
\section{The Phase Diagram of a Generic $V(\pdgp)$}
\vskip 5mm
\setcounter{section}{0}
\renewcommand{\thesection}{A}
In this Appendix we briefly review the necessary theoretical aspects of 
multi-cut phases of $\pdgp$. The first part of our discussion will 
apply for a generic polynomial $V(\pdgp)$. Then, in the second part of the 
Appendix, we will specialize to the sextic potential (\ref{sextic}). 

For practical reasons, we eliminate some (or all) of the couplings in 
$V(\pdgp)$ in terms of the end-points of the segments containing the 
eigenvalue distribution of 
$\pdgp$, and use the latter as (part of) the coordinates in the phase diagram.
In this way we can find rather easily which couplings in $V(\pdgp)$ are needed
to generate an eigenvalue distribution of $\pdgp$ with a prescribed set of 
support segments.

\subsection{A generic potential $V(\pdgp)$}

The saddle-point equation governing the Dyson gas of eigenvalues of 
$\pdgp$ is \cite{ambjorn, periwal}
\beq\label{dyson}
{\rm Re} F(s-i\epsilon) = {1\over 2} V'(s)\,. 
\eeq
By definition (see Eq. (\ref{FF}))
\beq\label{fw}
F(w) = {1\over N} \langle \trn {1\over w-\pdgp} \rangle = {1\over N} \sum_{i=1}^N \langle {1\over w-s_i}\rangle
\eeq
(where $s_i$ are the eigenvalues of $\pdgp$). Thus, as usual, 
\beq\label{realax}
F(s-i\epsilon) = {\rm P.P.}{1\over N} \sum_{j=1}^N \langle {1\over s-s_j}\rangle + i\pi\tilde\rho (s)\,,
\eeq
where $\tilde\rho(s)$ 
is the density of eigenvalues of $\pdgp$.

In order to study multi-cut configurations of $\tilde\rho(s)$, we also need the auxiliary function \cite{david} 
\beq\label{gs}
G(s) = \int\limits_{a_1>0}^s d\mu \left( V'(\mu) - 2F(\mu-i\epsilon)\right)\,.
\eeq
In the last equation $a_1$ is the lowest branch point of $F(w)$.
Thus, from (\ref{realax}) and (\ref{dyson}), we see that for $s$ real and in the support of eigenvalues,
\beq\label{sreal}
G(s) = - 2\pi i \int\limits_{a_1}^s \tilde\rho(\mu) d\mu
\eeq
is pure imaginary. $-\Im G(s) = 2\pi\tilde\rho(s)$ is then positive and monotonically increasing (and reaches $2\pi$ when $s$ hits 
the largest branch point).

{\it Stability of multi-cut distributions} ~~~ How do we know that a given distribution of eigenvalues is stable against migration
of eigenvalues from one place to another? To answer this question, consider the Dyson gas energy functional 
\beq\label{dysongas}
\sef = \int\limits_{s\geq 0} \tilde\rho(s) V(s) ds - {1\over 2} \int\limits_{s,\mu\geq 0} \tilde\rho(s) \tilde\rho(\mu) 
\log (s-\mu)^2 ~ds d\mu
\,.
\eeq
A general variation of (\ref{dysongas}) under $\tilde\rho(s)\rightarrow\tilde\rho(s) +\delta\tilde\rho(s) $ is 
\beq\label{variation}
\delta\sef = \int\limits_{s\geq 0} V(s) \delta\tilde\rho(s) ds - \int\limits_{s,\mu\geq 0} \tilde\rho(\mu) \log (s-\mu)^2 \delta
\tilde\rho(s) ~ds d\mu
\,.
\eeq
Moving an eigenvalue from $s_i$ to $s_f$ corresponds to $\delta\tilde\rho(s) = (1/N)[\delta(s-s_f) - \delta(s-s_i)]$. Thus, from 
(\ref{variation}) and (\ref{gs}) (and after some work) we can show that such a move costs 
\beq\label{cost}
\Delta\sef = {1\over N} \left[G(s_f) - G(s_i)\right]
\eeq
in energy \cite{david}. Such a rearrangement of eigenvalues costs energy only if 
\beq\label{cost1}
\Re\Delta\sef >0\,,
\eeq
and therefore (\ref{cost1}) is the stability condition against such a rearrangement. Thus, a multi-cut $F(w)$,
where the eigenvalues coalesce into $n$ segments $$[a_1, a_2]\bigcup [a_3, a_4]\bigcup \cdots \bigcup [a_{2n-1}, a_{2n}],$$ would be stable 
against migration of eigenvalues between neighboring cuts if and only if (\ref{cost1}) would hold for all neighboring pairs
of cuts, and in both directions. Since $G(s)$ is real on the segments on the real axis
that connect the cuts, this stability condition means 
\beq\label{global}
G(a_3) = G(a_2),\, G(a_5) = G(a_4), \cdots G(a_{2n-1}) = G(a_{2n-2})\,.
\eeq
In addition, of course, $\Re G(s)<0$ cannot happen anywhere for $s\geq 0$. 
The $n-1$ equations (\ref{global}) comprise the desired stability condition for such an eigenvalue distribution. 
In addition to these conditions, we have to make sure that along the cuts themselves $-\Im G(s)>0$, which is just the 
condition that $\tilde\rho(s)$ be positive.

The $n-1$ equations (\ref{global}), together with the obvious analytic properties of $F(w)$ and its asymptotic behavior 
\beq\label{asymptotic}
F(w)\sim {1\over w}
\eeq
as $w\rightarrow\infty$, determine $F(w)$ uniquely. Indeed, as is well known, for a 
generic $V(\pdgp)$, in view of (\ref{dyson}) and (\ref{asymptotic}) (and as we discussed at the end of Section 2), $F(w)$ 
(with $n$ cuts) must be of the form
\beq\label{fansatz}
F(w) = {1\over 2} V'(w) - P(w) \sqrt{\prod_{l=1}^{2n}(w-a_l)}\,,
\eeq
where 
\beq\label{p}
P(w) = {c_{-1}\over w} + \sum_{l=0}^{{\rm deg}\,V -n-1} c_l w^l\,.
\eeq
Here $a_1<a_2\cdots <a_{2n}$ and $c_{-1}\neq 0$ only if $a_1=0$ (see Section 2). If $a_1>0$, then $c_{-1} = 0$, and thus in such a 
case, there are $ 2n (a's) + ({\rm deg}\,V -n ) (c's) = n + {\rm deg}\,V $ 
independent parameters in the expression (\ref{fansatz}) for $F(w)$. On the other hand, there are ${\rm deg}\,V +1 $ conditions 
from the asymptotic behavior (\ref{asymptotic}) plus additional $n-1$ conditions from (\ref{global}), which comprise a total of 
${\rm deg}\,V + n$ conditions, equal to the number of unknown paramters. This balance remains if $c_{-1}$ appears in the game
as an unknown parameter, because then $a_1=0$, so that the number of parameters does not change. Finally, we have to remember to 
impose the positivity constraint 
\beq\label{positivity}
\tilde\rho(x) = {1\over\pi}\Im F(x - i\epsilon) \geq 0\,,
\eeq
which translates into a set of inequalities among the $a'$s and $c'$s.

{\it A convenient local parametrization of the phase diagram}~~~
Recall, that the phases of $\pdgp$ are specified by the number of segments in the support
of $\tilde\rho (s)$, i.e., the number of cuts in $F(w)$ (and whether these cuts have $w=0$ as a branch point or not.)
Thus, instead of the usual description of the phase structure in terms of the ${\rm deg}V$ couplings in $V(\pdgp)$, our strategy is 
to use ${\rm deg}V$ parameters out of the $2n$ branch-points $a_1, \cdots, a_{2n}$ of $F(w)$ and the ${\rm deg} P$ 
coefficients $c_k$ (with the total number ${\rm deg} V$ first saturated by the $a$'s in ascending order), which we refer to as 
``{\em phase coordinates}", to express (in a given phase) the couplings appearing in $V(\pdgp)$ (such as $m^2, \lambda $ and $ g$ in 
(\ref{sextic})), as well the as the $c_k$'s and $a_k$'s complementary to the phase coordinate parameter set. 
(See our discussion of the sextic potential below for concrete examples of this paramtrization.)

We have to be careful in giving the expressions for, say, the couplings of $V$, in terms of the phase coordinates. This because 
for a given configuration of $F(w)$, the equations from which we are to eliminate the couplings of $V$ (such as the 
triad  $m^2, \lambda $ and  $g$ in (\ref{sextic})) as functions of the phase coordinates may have several 
solutions (in other words, the  couplings in $V$ are generally multivalued functions of the phase coordinates in a given phase).
Thus, in a given phase, we must of course choose the parametrization of couplings in $V$ which yields the 
minimal $\sef$ appropriate for that phase.

This alternative 
parametrization is more convenient for our purposes in Section 5. Indeed, 
once we are successful in expressing the couplings 
in $V$ as functions of the phase coordinates, it will be very easy 
for us to tune the couplings in $V(\pdgp)$ to a generic point in a given phase and also to approach the phase 
boundaries in a controlled manner. In particular, phase transitions appear here, for example, when 
branch points collide and become equal (at some common real positive value $a$). This process removes 2 a's and thus closes 
one cut $(n\rightarrow n-1)$ but adds an additional term to $P(w)$. The number of unknown parameters drops by 1, but so does the number of 
stability conditions (\ref{global}). In the other dircetion, we can obviously reach the same coexistence point, by 
tuning the parameters of $P(w)$ to a point where it develops a linear factor 
$(w-a) = \sqrt{(w-a)^2}$ (with $a\geq 0$). 
Obviously, when these alternative phase coordinates approach a point on the coexistence surface from two different sides of the 
pahse transition, the respective sets of couplings of $V$, expressed as sets of functions of the two phase coordinate patches, coincide. 
Thus, they lead to the same $\sef$, which means that such a point is indeed a point on the phase boundary.

\subsection{Results for the sextic potential}
From (\ref{fansatz}) and (\ref{sextic}), the general form of $F(w)$ is 
\beq\label{fsextic}
F(w) = {1\over 2}\left(m^2  +\lambda w + g w^2\right) -P(w)\sqrt{{\rm polynomial}}\,.
\eeq
{\bf A. single cut, disk phase}: ~~~
Here
\beq\label{1cutdisk}
F(w) = {1\over 2}\left(m^2  +\lambda w + g w^2\right) -
\left({s\over w} + t + uw\right)\sqrt{w(w-a)}\,.
\eeq
There is a single cut, so (\ref{global}) is trivial in this case, and (\ref{positivity}) holds manifestly. We need only impose
(\ref{asymptotic}). In the end, we find
\beqra\label{parameters1cutdisk}
&& u= {16 - 8as -2a^2 t\over a^3}\nonumber\\
&& g = 2u\nonumber\\
&& \lambda = 2t - au = {4a^2t + 8as -16\over a^2}\quad {\rm and}\nonumber\\
&& m^2 = 2s - at - {a^2u\over 4} = {8as -a^2t -8\over 2a}\,.
\eeqra
The phase coordinates are $a, t$ and $s$.

{\bf B. single cut, annular phase}: ~~~
We have 
\beq\label{1cutannulus}
F(w) = {1\over 2}\left(m^2  +\lambda w + g w^2\right) -
\left( t + uw\right)\sqrt{(w-a)(w-b)}\,.
\eeq
Here $0<a<b$. Again, there is a single cut, so (\ref{global}) is trivial in this case too, and also (\ref{positivity}) holds manifestly.
We need 
only impose (\ref{asymptotic}). In the end, we find
\beqra\label{parameters1cutannulus}
&& u= {16 -2t(a-b)^2\over (a+b)(a-b)^2}\nonumber\\
&& g = 2u\nonumber\\
&& \lambda = 2t - u(a+b)  = 4t -{16\over (a-b)^2}\quad {\rm and}\nonumber\\
&& m^2 = -{(a-b)^2\over 4}u - t(a+b) = -{4\over a+b} -
t { a^2 + 6ab + b^2\over 2(a+b)}\,.
\eeqra
The phase coordinates are $a, b$ and $t$.  

{\bf C. two cuts, disk phase}: ~~~
We have
\beq\label{2cutdisk}
F(w) = {1\over 2}\left(m^2  +\lambda w + g w^2\right) -
\left( {s\over w} + t \right)\sqrt{w(w-a)(w-b)(w-c)}\,.
\eeq
Here $0<a<b<c$. Note that in this case we can trade the three couplings $m^2, \lambda$ and $g$ for the 
three branch points $a,b$ and $c$. In this case there are two cuts, so for the first 
time (\ref{global}) is not trivial. We first impose (\ref{asymptotic}). We find
\beqra\label{parameters2cutdisk}
&& s= {8\over a^2 + b^2 + c^2 -2(ab + ac + bc)}\nonumber\\
&-&{1\over 2} {a^3 + b^3 + c^3 - a^2(b+c) - b^2(a+c) - c^2(a+b) + 2abc\over a^2 + b^2 + c^2 - 2(ab+ac+bc)}t\nonumber\\
&& g = 2t\nonumber\\
&& \lambda = 2s - t(a+b+c)\quad {\rm and}\nonumber\\
&& m^2 = -(a+b+c)s - {a^2 + b^2 + c^2 -2(ab+ac+bc)\over 4}t\,.
\eeqra
We have yet to impose (\ref{global}), which is why $t$ was not eliminated yet. 
Before doing that, we impose (\ref{positivity}). 
Our conventions are always to take each cut from the appropriate branch point to the 
left on the real axis. Thus, after some work, we find from (\ref{2cutdisk}) 
\beqra\label{density}
\pi\tilde\rho(x) &=& \Im F(x-i\epsilon)=\nonumber\\{}\nonumber\\
&&\left\{\begin{array}{c} 
-({s\over x} + t)\sqrt{x(a-x)(b-x)(c-x)}\quad\quad 0<x<a\\{}\\
+({s\over x} + t)\sqrt{x(x-a)(x-b)(c-x)}\quad\quad b<x<c\\{}\\
0\quad\quad {\rm otherwise}\,.\end{array}\right.
\eeqra
We have to impose (\ref{positivity}) on (\ref{density}). This means 
\beqast
{s\over x} + t <0\quad\quad {\rm for }\quad 0<x<a\nonumber\\{\rm and}\\
{s\over x} + t >0\quad\quad {\rm for }\quad b<x<c\,.
\eeqast
Thus, we must have\footnote{It is straightforward to check that the 
following inequalities hold for the ensembles corresponding to 
Figures 4 and 5 in Section 5.}  
\beq\label{inequalities}
t>0\quad\quad {\rm and}\quad\quad -bt\leq s\leq -at<0\,,
\eeq
where $s$ is given in (\ref{parameters2cutdisk}).
We are now ready to impose (\ref{global}). Here it simply means $G(a)=G(b)$, namely,
\beq\label{globaldisk}
\int\limits_a^b ({s\over x} + t)\sqrt{x(x-a)(b-x)(c-x)}\,dx = 0\,.
\eeq
Note from (\ref{inequalities}) that $-b<s/t<-a$, and thus the factor multiplying the square root in (\ref{globaldisk}) flips its sign in 
the integration domain, so that the integral on the LHS of (\ref{globaldisk}) may vanish. The latter equation may be expressed in terms
of the elliptic integrals 
\beqra\label{elliptic}
I(a,b,c) &=& \int\limits_a^b \sqrt{x(x-a)(x-b)(x-c)} dx \nonumber\\{\rm and}\nonumber\\
J(a,b,c) &=& \int\limits_a^b \sqrt{x(x-a)(x-b)(x-c)} {dx\over x}\,.
\eeqra
Following the usual procedure, we may express $I$ and $J$ in terms 
of complete elliptic integrals in a 
straightforward manner. (We do not bother to write these expressions here, 
since for our purposes in Section 5 we evaluated $I$ and $J$ numerically.)

Finally, substituting these expressions in (\ref{globaldisk}) we obtain 
\beq\label{t}
s(t,a,b,c)J(a,b,c)+tI(a,b,c)=0
\eeq
which we solve for $t$ (recall from (\ref{parameters2cutdisk}) that $s$ is merely 
linear in $t$, and also that $t>0$, in view of (\ref{inequalities}).) 
Once $t(a,b,c)$ is known, we can go back to (\ref{parameters2cutdisk}) and evaluate
$m^2(a,b,c), \lambda(a,b,c)$ and $g(a,b,c)$ explicitly.  Our phase coordinates in this case are thus $a, b$ and $c$.


\vskip 20mm
\begin{center}
{\bf ACKNOWLEDGEMENTS}
\end{center}
We thanks H. Orland for clarifying the Abstract. 
The work of A.Z. was supported in part by the National Science Foundation 
under Grant No. NSF-PHY99-07949. RTS acknowledges M. Papas and 
support from NSF-DMR-9985978 and also the ITP at UC Santa-Barbara for its 
hospitality, where part of this work was done. JF's research has been 
supported in part by the Israeli Science Foundation grant number 307/98 
(090-903). 


\pagebreak

\begin{figure}[h]
\centering{\epsfig{file=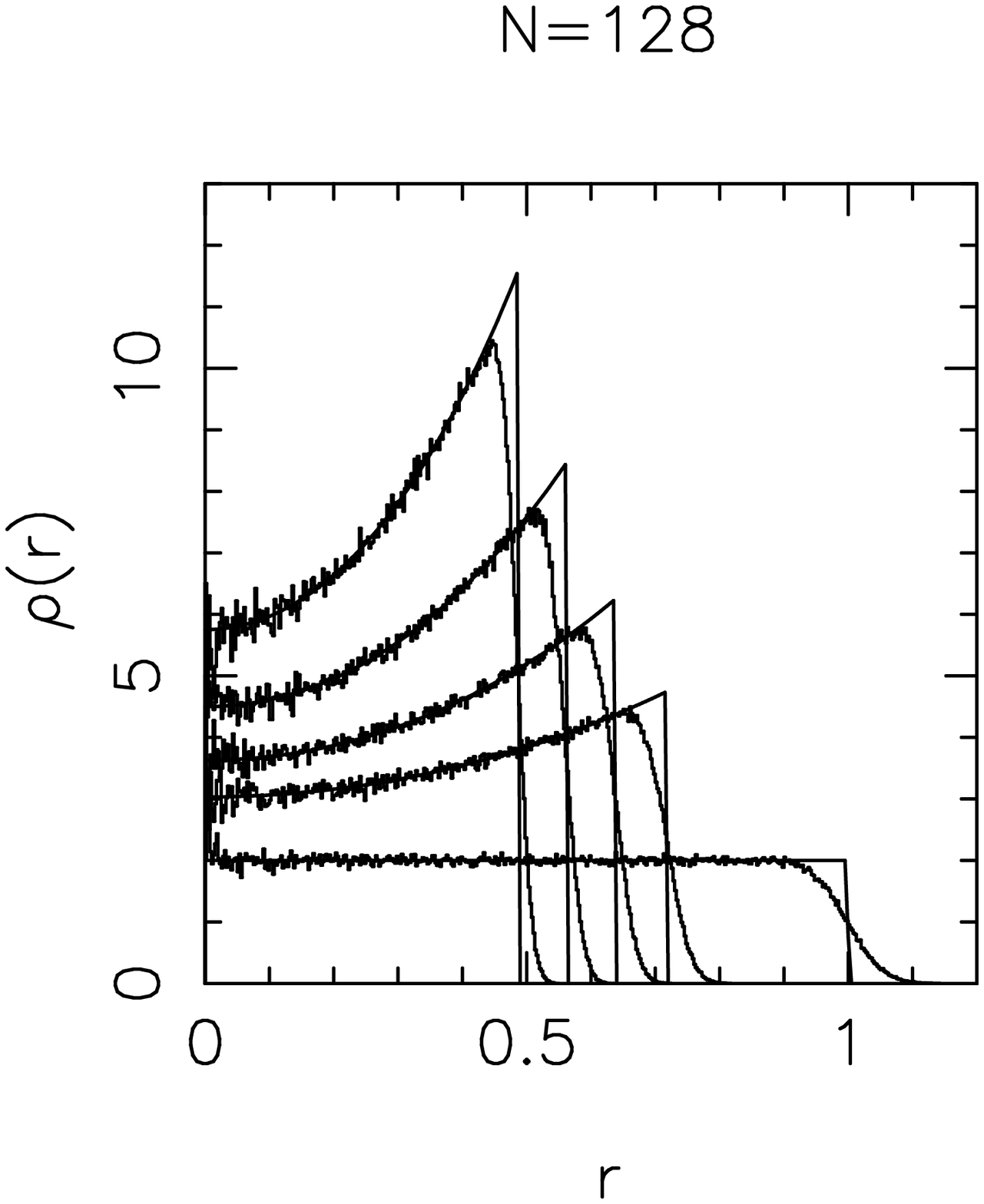,width=4in,height=4in}}
\caption{Comparison between Monte-Carlo measurements of the density of 
eigenvalues $\rho (r)$ of matrices $\phi$ of size 128x128, taken from the 
quartic ensemble $V(\pdgp) = 2m^2\pdgp + g(\pdgp)^2 $ with $m^2=0.5$ 
(disk phase) and for $g = 0, 0.5, 1, 2, 4$ ($g$ increases from bottom to 
top), compared to the analytical results of \cite{fz1} (solid lines). 
At $g=0$ we obtain Ginibre's Gaussian ensemble with $V=\pdgp$, with its unit disk of eigenvalues.}
\end{figure}

\begin{figure}[h]
\centering{\epsfig{file=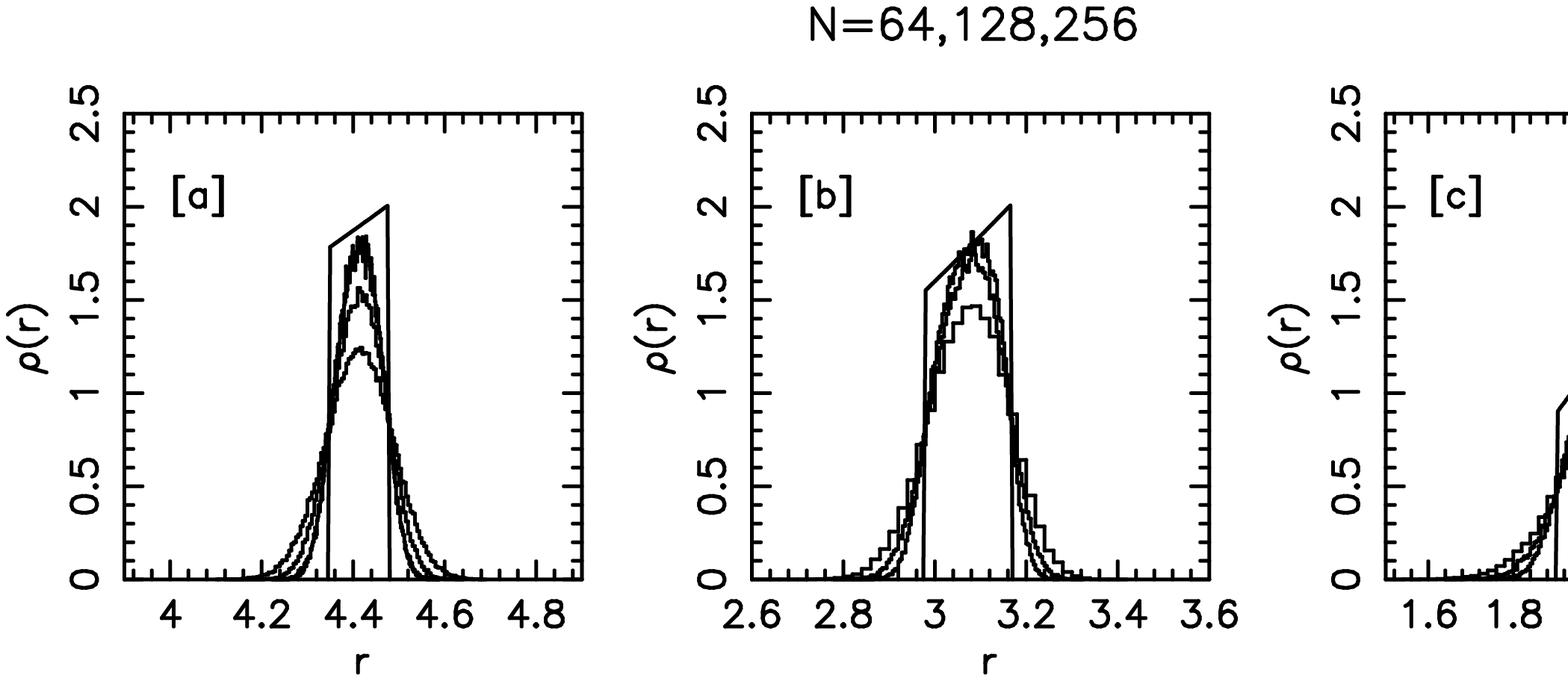,width=2.5in,height=2.5in}}
\vskip -2cm
\caption{Results of Monte-Carlo measurements of the density of 
eigenvalues $\rho (r)$ of matrices $\phi$ of sizes corresponding to 
$N=64, 128$ and $256$, taken from the quartic ensemble with 
$m^2=-\mu^2 =-0.5$ (annular phase) for various values of the quartic 
coupling: $g=0.025$ in [a], $g=0.05$ in [b] and $g=0.1$ in [c]. These are 
compared to the analytical results of 
\cite{fz1} (solid lines). As $N$ increases, the numerical results 
converge monotonically to the analytical results.}
\end{figure}

\begin{figure}[h]
\centering{\epsfig{file=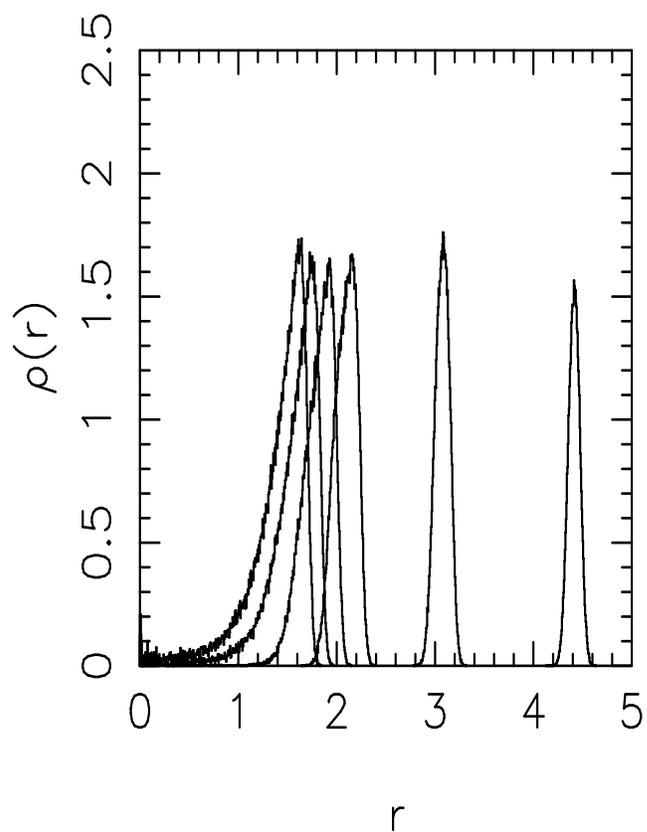,width=4in,height=4in}}
\caption{Monte-Carlo measurements of the density of eigenvalues 
$\rho (r)$ of matrices $\phi$ of size 128x128, taken from the 
the quartic ensemble with $\mu^2=0.5$ and for 
$g = 0.025, 0.05, 0.1, 0.125, 0.15$ and $0.175$ ($g$ increases from 
right to left). The first three profiles on the right (corresponding to the 
three lowest values of $g$) evidently belong to the annular phase. The fourth 
density profile from the right is the critical one 
(corresponding to $g_c=0.125$). Finally, The last two profiles 
(which correspond to the two higher values of $g$) belong to the 
disk configuration.}
\end{figure}

\begin{figure}[h]
\centering{\epsfig{file=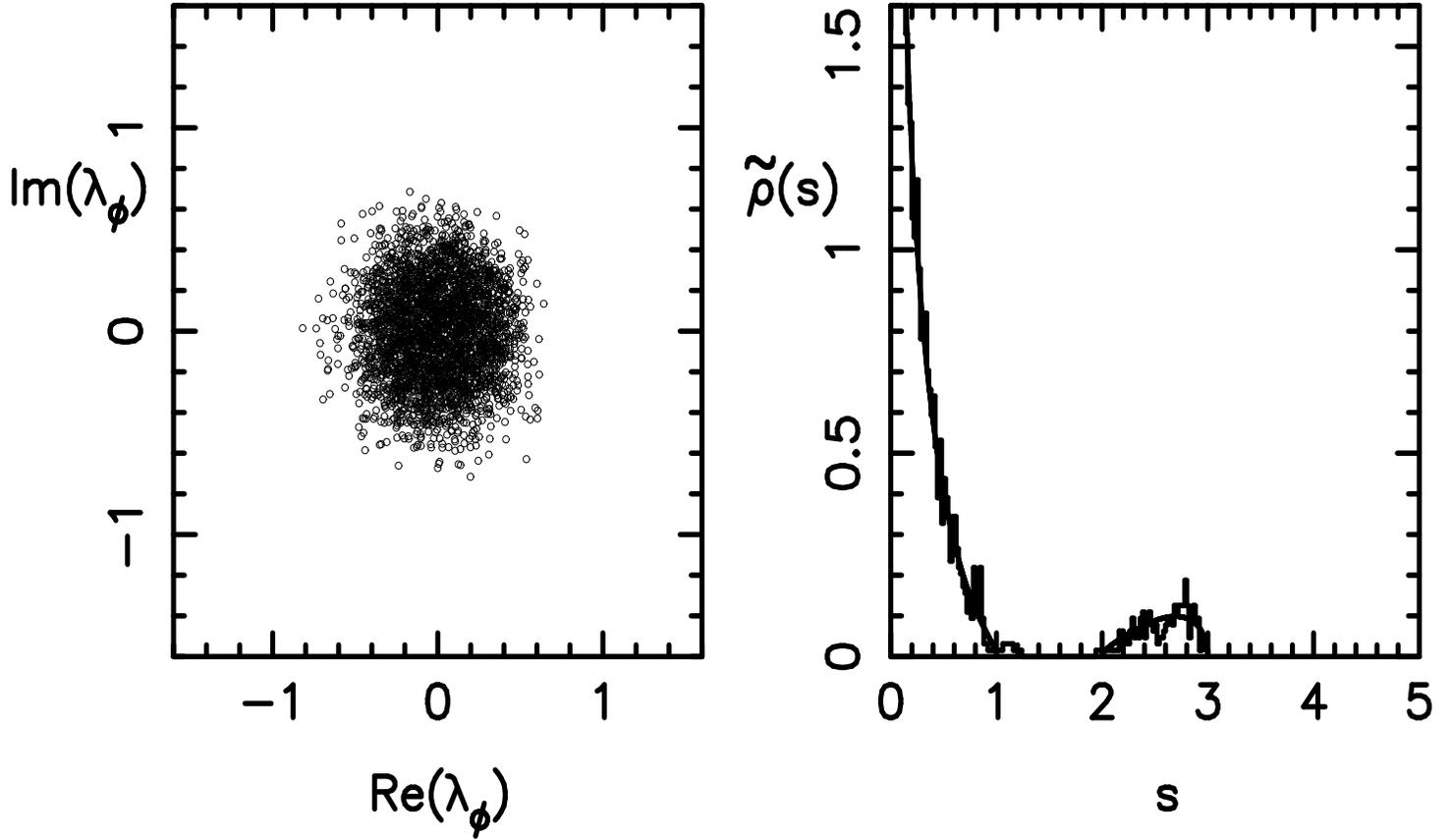,width=4in,height=4in}}
\caption{Scatter plot of the eigenvalues of 
matrices $\phi$ of size 32x32, taken from (\ref{sextic}) with 
$m^2=7.372, \lambda=-6.116 $ and $g=1.372$ (left), and the corresponding 
density of eigenvalues $\tilde\rho(s)$ of $\pdgp$ (right). The solid line on 
the right is the analytical curve corresponding to (\ref{density}). 
The support of $\tilde\rho(s)$ is split into the two segments 
$\{0\leq s \leq a=1\}\bigcup\{b=2\leq s \leq c=3\}$, while the support of 
$\rho(r)$ on the left is manifestly a disk.}
\end{figure}

\begin{figure}[h]
\centering{\epsfig{file=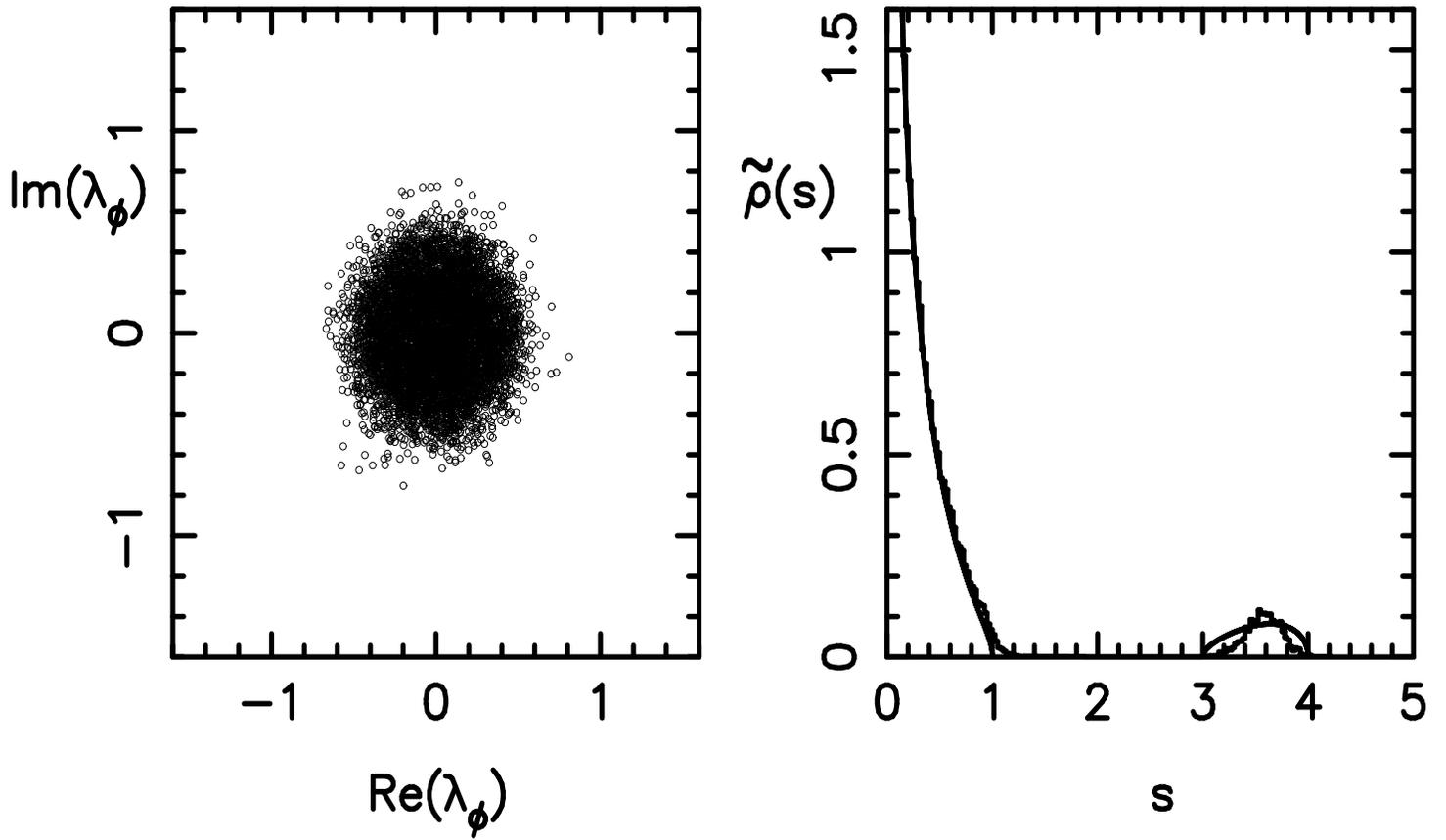,width=4in,height=4in}}
\caption{Similarly to Figure 4, but with 
$m^2=6.403, \lambda=4.184$ and $g=0.713$. 
The support of $\tilde\rho(s)$ is split into the two segments 
$\{0\leq s \leq a=1\}\bigcup\{b=3\leq s \leq c=4\}$, while the support of 
$\rho(r)$ on the left remains a disk.}
\end{figure}

\end{document}